\documentclass[aps,prl,twocolumn,amsmath,amssymb,superscriptaddress]{revtex4-1}

\usepackage{blindtext}

\usepackage{graphicx}
\usepackage{amssymb}
\usepackage{color}
\usepackage{psfrag}
\usepackage{ifsym}
\usepackage{epstopdf}
\usepackage{soul}
\usepackage{tabularx}
\usepackage{hyperref}

\usepackage{blindtext}
\usepackage{graphicx}
\usepackage{amssymb}
\usepackage{color}
\usepackage{psfrag}
\usepackage{ifsym}
\usepackage{epstopdf}
\usepackage{soul}
\usepackage{tabularx}

\usepackage{epsfig,color}
\usepackage{amsmath,amssymb}

\usepackage{hyperref}

\usepackage{float}

\begin{document}
\title{Scalable performance in solid-state single-photon sources}

\author{J.~C.~Loredo}\email{Corresponding author: juan.loredo1@gmail.com}
\affiliation{Centre for Engineered Quantum Systems, Centre for Quantum Computation and Communication Technology, School of Mathematics and Physics, University of Queensland, Brisbane, Queensland 4072, Australia}
\author{N.~A.~Zakaria}
\affiliation{Centre for Engineered Quantum Systems, Centre for Quantum Computation and Communication Technology, School of Mathematics and Physics, University of Queensland, Brisbane, Queensland 4072, Australia}
\author{N.~Somaschi}
\affiliation{CNRS-LPN Laboratoire de Photonique et de Nanostructures, Universit\'e Paris-Saclay, Route de Nozay, 91460 Marcoussis, France}
\author{C.~Anton}
\affiliation{CNRS-LPN Laboratoire de Photonique et de Nanostructures, Universit\'e Paris-Saclay, Route de Nozay, 91460 Marcoussis, France}
\author{L.~De~Santis}
\affiliation{CNRS-LPN Laboratoire de Photonique et de Nanostructures, Universit\'e Paris-Saclay, Route de Nozay, 91460 Marcoussis, France}
\affiliation{Universit\'e Paris-Sud, Universit\'e Paris-Saclay, F-91405 Orsay, France}
\author{V.~Giesz}
\affiliation{CNRS-LPN Laboratoire de Photonique et de Nanostructures, Universit\'e Paris-Saclay, Route de Nozay, 91460 Marcoussis, France}
\author{T.~Grange}
\affiliation{Universit\'e Grenoble-Alpes, CNRS, Institut N\'{e}el, ``Nanophysique et semiconducteurs'' group, F-38000 Grenoble, France}
\author{M.~A.~Broome}
\affiliation{Centre for Engineered Quantum Systems, Centre for Quantum Computation and Communication Technology, School of Mathematics and Physics, University of Queensland, Brisbane, Queensland 4072, Australia}
\affiliation{Present address: Centre of Excellence for Quantum Computation and Communication Technology, School of Physics, University of New South Wales, Sydney, New South Wales 2052, Australia}
\author{O.~Gazzano}
\affiliation{CNRS-LPN Laboratoire de Photonique et de Nanostructures, Universit\'e Paris-Saclay, Route de Nozay, 91460 Marcoussis, France}
\affiliation{Present address: Joint Quantum Institute, National Institute of Standards and Technology, University of Maryland, Gaithersburg, MD, USA}
\author{G.~Coppola}
\affiliation{CNRS-LPN Laboratoire de Photonique et de Nanostructures, Universit\'e Paris-Saclay, Route de Nozay, 91460 Marcoussis, France}
\author{I.~Sagnes}
\affiliation{CNRS-LPN Laboratoire de Photonique et de Nanostructures, Universit\'e Paris-Saclay, Route de Nozay, 91460 Marcoussis, France}
\author{A.~Lemaitre}
\affiliation{CNRS-LPN Laboratoire de Photonique et de Nanostructures, Universit\'e Paris-Saclay, Route de Nozay, 91460 Marcoussis, France}
\author{A.~Auffeves}
\affiliation{Universit\'e Grenoble-Alpes, CNRS, Institut N\'{e}el, ``Nanophysique et semiconducteurs'' group, F-38000 Grenoble, France}
\author{P.~Senellart}
\affiliation{CNRS-LPN Laboratoire de Photonique et de Nanostructures, Universit\'e Paris-Saclay, Route de Nozay, 91460 Marcoussis, France}
\affiliation{D\'epartement de Physique, Ecole Polytechnique, Universit\'e Paris-Saclay, F-91128 Palaiseau, France}
\author{M.~P.~Almeida}
\affiliation{Centre for Engineered Quantum Systems, Centre for Quantum Computation and Communication Technology, School of Mathematics and Physics, University of Queensland, Brisbane, Queensland 4072, Australia}
\author{A.~G.~White}
\affiliation{Centre for Engineered Quantum Systems, Centre for Quantum Computation and Communication Technology, School of Mathematics and Physics, University of Queensland, Brisbane, Queensland 4072, Australia}

\begin{abstract}
The desiderata for an ideal photon source are high brightness, high single-photon purity, and high indistinguishability. Defining brightness at the first collection lens, these properties have been simultaneously demonstrated with solid-state sources, however absolute source efficiencies remain close to the $1\%$ level, and indistinguishability only demonstrated for photons emitted consecutively on the few nanosecond scale. {Here we employ deterministic quantum dot-micropillar devices to demonstrate solid-state single-photon sources with scalable performance. In one device, an absolute brightness at the output of a single-mode fibre of $14\%$ and purities of $97.1$--$99.0\%$ are demonstrated. When non-resontantly excited, it emits a long stream of photons that exhibit indistinguishability up to $70\%$---above the classical limit of $50\%$---even after 33 consecutively emitted photons, a $400$~ns separation between them. Resonant excitation in other devices results in near-optimal indistinguishability values: $96\%$ at short timescales, remaining at $88\%$ in timescales as large as $463$~ns, after $39$ emitted photons. The performance attained by our devices brings solid-state sources into a regime suitable for scalable implementations.}
\end{abstract}

\maketitle

\noindent Photon indistinguishability---responsible for unique quantum phenomena with no classical counterpart, notably photon bunching via interference~\cite{HOM:Mandel}---has been demonstrated in various physical systems~\cite{indAtom:Kuhn,indIonR:Monroe,indNV:Hanson,indMol:Zumbusch,indSCirc:Wallraff,indSPDC:Franson,indQD:Yamamoto,indQD:Senellart}, resulting in a broad range of applications in photonic quantum technologies~\cite{QuantTech:Obrien}, including quantum teleportation~\cite{qtel:Zeilinger,qtelM:Pan}, generation of entangled photon sources~\cite{entSource:Kwiat,entSource:Wong,entSource:Senellart}, and linear-optics quantum computation~\cite{KLM:KLM,LOQC:Milburn}. However, achieving conclusive indistinguishability, i..e. above $50\%$ (the classical limit), while simultaneously displaying high single-photon purity and high absolute brightness is experimentally challenging.

Semiconductor quantum dots (QDs) inserted in photonic structures~\cite{spDevice:Imamoglu,spEnhEmis:Yamamoto,spQDEff:Yamamoto,spQDPhCr:Lodahl,spQDEffOut:Lodahl} are a rapidly improving technology for generating bright sources of indistinguishable single-photons. Addressing the excited states of the quantum dot using a {non}-resonant scheme early showed two-photon interference visibilities in the $70\%{-}80\%$ range~\cite{indQD:Yamamoto}, yet with limited collection efficiencies. Improvements in the efficiency have been made by deterministically placing the quantum dot in the centre of a photonic micro-cavity. Here the acceleration of photon emission into well defined cavity modes~\cite{enhancedEm:JMGerard}, due to Purcell enhancement, has enabled two-photon interference visibilities in the same range, with simultaneous efficiencies at the first collection lens around $80\%$~\cite{indQD:Senellart}. {Near-unity indistinguishability, in turn, has been achieved in recent years under strictly-resonant excitation of the quantum dot~\cite{indQDUnity:Pan,indQD995:Pan,10000:JWP},} whereas the recent development of electric control on deterministically coupled devices~\cite{spDetElecCont:Senellart}---thus with scalable fabrication---has now enabled strictly-resonant excitation in combination with Purcell enhancement, resulting in near-optimal single-photon sources~\cite{nearOpt:Senellart} with visibilities reaching the $99\%$ mark, simultaneous state-of-the-art extraction efficiency of $65\%$ and polarised brightness at the first lens around~$16\%$.

Albeit impressive, the reported efficiencies in these demonstrations are defined at the first lens, and poor optical collection results in low photon count rates available in practice. Consequently, absolute brightnesses remain around the $1\%$ mark, too low for practical scalable applications~\cite{QuantTech:Obrien}. In addition, direct measurements of indistinguishability via two-photon interference, so far, only employed photons consecutively emitted with a few nanosecond separation, {while a key question regarding the scalable potential of the developed sources is to determine how many consecutive photons exhibit high indistinguishability. A recent work obtained on quantum dots in microlenses reported a 40~\% drop in the indistinguishability over 10~ns only~\cite{QD:Thoma}.}

{In the present work, we demonstrate high absolute brightness and generation of indistinguishable  photons consecutively emitted over $463$~ns. Our measurements were performed on various quantum dot-micropillar devices, all obtained using a deterministic---thus scalable--- technology. Using a simple micropillar  (\emph{Device~$1$})~\cite{indQD:Senellart}, we demonstrate a high-purity single-photon source with an absolute brightness of $14\%$. That is, about one in seven laser pulses creates a high-purity single-photon at the output of a single-mode fibre. We also demonstrate robust and conclusive quantum interference between consecutively emitted photon pulses up to a first and thirty-third, separated by $400$~ns. Interference visibilities, under non-resonant excitation, reach maximum values of $70\%$ in short timescales, decreasing to plateaus above $60\%$ at longer temporal separations, and remain above the classical limit of $50\%$ even at high pump-powers. Using electrically controlled pillar devices~\cite{nearOpt:Senellart}  (\emph{Device~}$2$ and $3$) we demonstrate, under strictly resonant-excitation, indistinguishability reaching near-optimal values: $96\%$ at short timescales, remaining above $88\%$ at $463$~ns separation.}

\emph{Device~$1$} contains self-assembled InGaAs QDs grown by molecular beam epitaxy, positioned in between two layers of GaAs/AlAs distributed Bragg reflectors, consisting of $16$ ($36$) pairs acting as a top (bottom) mirror. Note that \emph{Device~$1$} is a pillar from the same batch as in Ref.~\cite{indQD:Senellart}.~Low-temperature \emph{in situ} lithography~\cite{inSituLith:Senellart} was employed to fabricate micropillars centred around a single QD with $50$~nm accuracy. The sample is mounted on a closed-cycle cryostat and is optically pumped by $5$~ps laser pulses at $80$~MHz repetition rate with wavelength tuned to $905.3$~nm, corresponding to one of the quantum dot excited states in its p-shell. We optimised our collection efficiency by judicious choice of optical elements, achieving an efficiency budget as follows. After emission from the micropillar, single-photons travel across the following elements, with measured transmittances $\eta_{elem}$, before reaching detectors: two cryostat windows with $\eta_{cryo}{=}(96\pm1)\%$; a microscope objective (Olympus LMPLN$10$XIR) with N.A.${=}0.3$ and $\eta_{obj}{=}(91\pm1)\%$; a dichroic mirror (Alluxa filters) used to separate single-photons from the laser path, with a measured attenuation at $905$~nm bounded to $>60$~dB extinction, while no appreciable loss is recorded at wavelengths corresponding to single-photon emission, we thus consider $\eta_{dich}{=}1$; $6$ mirrors and $2$ lenses, with an overall transmission of $\eta_{ml}{=}(95\pm1)\%$; and a $0.85$~nm FWHM band-pass filter (Alluxa filters) with $\eta_{bp}{=}(91\pm1)\%$ used to ensure that any residual scattered laser light is filtered out. Remaining losses are due to coupling to a single-mode fibre, where we estimate a fibre-coupling efficiency of $\eta_{fc}{=}(65\pm4)\%$, by comparing collection with a multimode fibre assumed to have a unity coupling efficiency. This results in an overall transmission of our optical setup of $\eta_{setup}{=}(49\pm3)\%$.
\begin{figure}[htp]
\centering
\includegraphics[width=.85\linewidth]{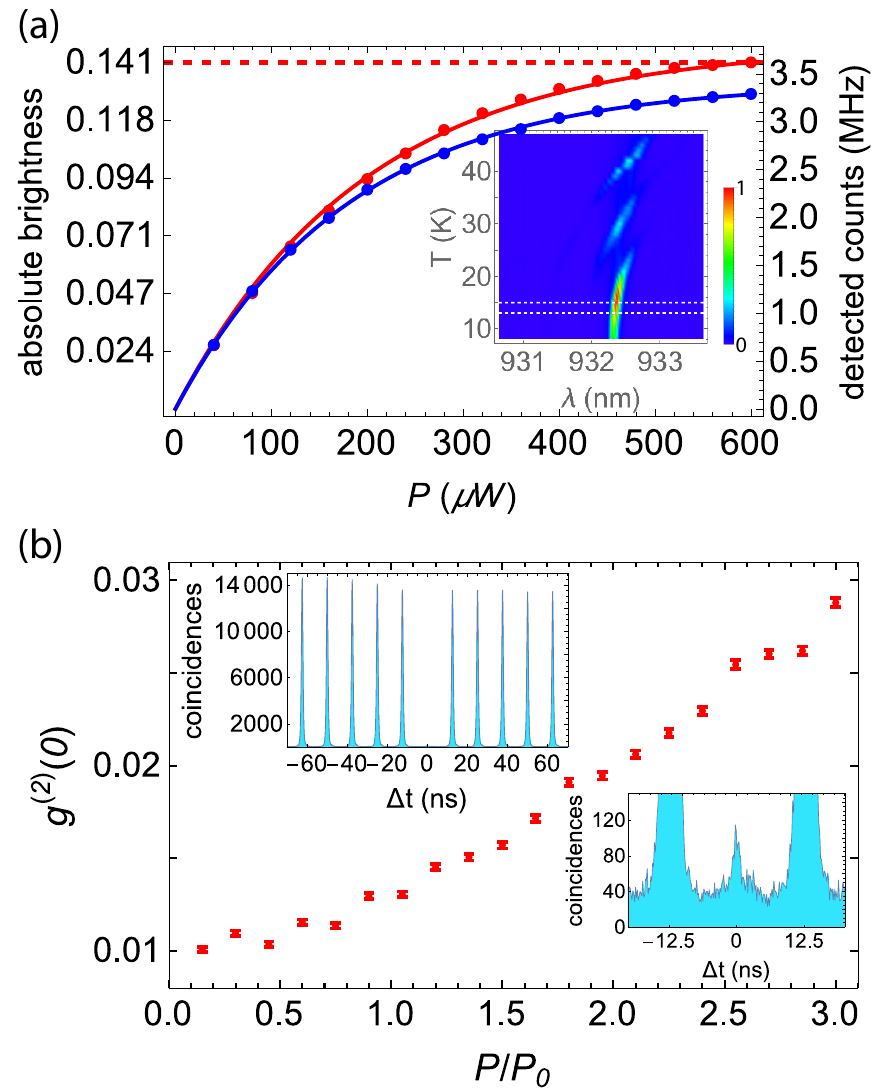}
\caption{Absolute brightness and purity of \emph{Device~$1$}.  a) Detected count rates at $T{=}15$~K (red), with the QD in resonance with the cavity mode, and $13$~K (blue), with the QD slightly detuned from the cavity. Solid curves represent fits to $R_0\left(1-\exp{(-P{/}P_0)}\right)$, with $P_0{=}197$~$\mu$W, and $R_0{=}3.8$~MHz for $T{=}15$~K, and $R_0{=}3.4$~MHz for $T{=}13$~K. Inset: QD spectra with varying temperature. b) Power-dependent $g^{(2)}(0)$ at T{=}$15$~K. Note that even three times above the saturation pump power the photon purity remians $>97$\%. Top inset shows the autocorrelation measurement for $P{=}1P_0$, and bottom inset zooms into the zero delay resolving the non-zero $g^{(2)}(0)$ from experimental noise.}
\label{fig:1}
\end{figure}

We characterise this device in terms of absolute brightness and purity, see Fig.~\ref{fig:1}. We detect large count-rates in a silicon avalanche photodiode (APD), as shown in the saturation measurements in Fig.~\ref{fig:1}a. The saturation curves are fitted to $R_0\left(1-\exp{(-P{/}P_0)}\right)$, where $R_0$ is an asymptotic rate value, and $P_0$ is the saturation power. The inset figure shows \emph{Device~1} spectra with varying temperature $T$. {The energy of the QD transition varies like the band gap of the semiconductor with temperature~\cite{Tscan:Pavesi}, whereas the cavity mode energy follows the temperature variation of the refractive index. Adjusting the temperature thus allows tuning the QD-cavity resonance. For the measurements presented in Fig.~\ref{fig:1}, the neutral exciton line is brought in resonance at $T{=}15$~K. The count-rates in pulsed configuration reach values as high as $3.6$~MHz.} In fact, for this measurement a known loss must be introduced in the optical path in order to properly quantify the available count-rates, as they are beyond the APD's (Perkin-Elmer SPCM{-}AQR{-}$14${-}FC) linear regime. This allows us to accumulate a high amount of statistics with notably short integration times. For instance, the inset in Fig.~\ref{fig:1}b shows a $g^{(2)}(\Delta t)$ measurement---second-order autocorrelation function with $g^{(2)}(0){=}0$ corresponding to an ideal single-photon state---at $P{=}P_0$, yielding a value of $g^{(2)}(0){=}0.0130\pm0.0002$, where the small error is reached with an integration time of only $29$ seconds. We in fact used about half the available counts after selecting one linear polarisation emitted by our device. Thus, in our setup, the same amount of statistics is achieved four times faster when the polariser is removed. Remarkably, we observe low multi-photon emission at all pump-powers, with a measured maximum value of $g^{(2)}(0){=}0.0288\pm0.0002$ at $P{=}3P_{0}$. We thus observe a single-photon purity $1{-}g^{(2)}(0)$ above $97\%$ even at maximum brightness. These values were extracted from integrating raw counts in a $2$~ns window---sufficiently larger than the $<0.5$~ns lifetime~\cite{indQD:Senellart}---around the peak at zero delay compared to the average of the 10 adjacent lateral peaks, {without any background subtraction}. Error bars in this work are deduced from assuming poissonian statistics in detected events.

Our APD efficiency of $32\%$---measured using the approach of Ref.~\cite{detEff:Hadfield}---$80$~MHz pump rate, and $3.6$~MHz detected count rate corresponds to an absolute brightness---the probability-per-laser-pulse of finding a spectrally-isolated high-purity single-photon at the output of a single-mode fibre---of $14\%$, the highest reported to date.
{Such absolute brightness represents a clear improvement over what has been previously achieved with quantum dot-based photon sources. For instance, a drastic contrast between performance at the first lens and actual detected count rates has been common until now, e.g., reporting a brightness as high as $72\%$ while detecting $65$~kHz~\cite{spQDHighEff:JMGerard}, or $143$~MHz collected on the first lens but only $72$~kHz available on detection~\cite{143MHz:Reitzenstein}. Detected rates of $4.0$~MHz at the single-photon level have been reported~\cite{spQDHighFreq:Bouwm}, however without coupling into a single-mode fibre and at the cost of high multi-photon contribution with $g^{(2)}(0){=}0.4$. In fact, our source greatly exceeds, in terms of absolute brightness, the performance of any other single-photon source from any physical system, including the well established Spontaneous Parametric DownConversion source---so far considered as the premier photon source---where the equivalent (triggered) absolute brightness is well below $1\%$.}
\begin{figure}[h!]
\begin{center}
\includegraphics[width=.85\linewidth]{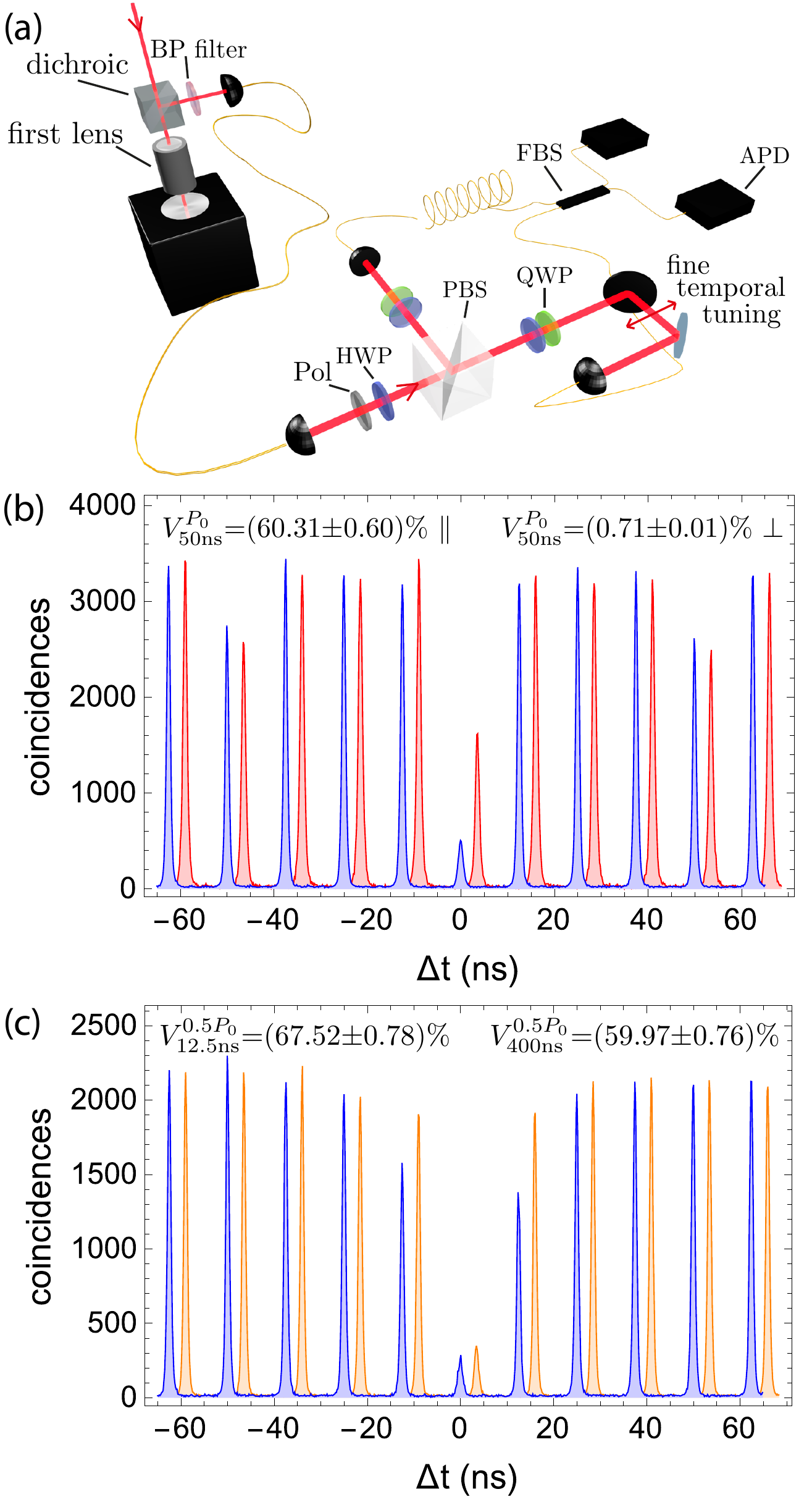}
\caption{Two-photon interference between temporally-distant photons. a) A simple unbalanced Mach-Zehnder interferometer with a path-length difference of $\Delta\tau_e$ probes the indistinguishability of two photons emitted with the same $\Delta\tau_e$ temporal separation. b) Interference histograms of orthogonal- (red) and parallel-polarised (blue) photons with $\Delta\tau_e{=}50$~ns, at saturation of the quantum dot. (Note the suppression at $\Delta\tau_e$, see text for details). c) Interference of parallel-polarised photons with $\Delta\tau_e{=}12.5$~ns (blue) and $\Delta\tau_e{=}400$~ns (orange), taken at $P{=}0.5P_0$. A temporal offset of $3.5$~ns has been introduced between histograms for clarity.}
\label{fig:2}
\end{center}
\end{figure}

{We note that, given our setup collection efficiency of $\eta_{setup}{=}49\%$, \emph{Device~$1$} exhibits---for the neutral exciton state---a brightness at the first lens of $29\%$. Deducing the exciton lifetime from the correlation curves at low excitation power, we estimate the Purcell factor of the device to be around $F_p{=}2$, and the fraction of emission into the cavity mode around $66\%$. Considering an output coupling efficiency of $90\%$, the measured brightness in the first lens could reach $60\%$ with a unity probability to find the QD in the neutral exciton state. However, as evidenced in the inset of Fig.~\ref{fig:1}a, the present QD also presents an non-negligible probability to emit from the positively- or negatively-charged exciton transition that are brought in resonance at higher temperatures. As a result, the probability of the quantum dot to be in the neutral exciton is reduced leading to the measured $29\%$ brightness at the first lens. Note that this instability of the charge state was not observed originally in the devices under study, see Ref.~\cite{indQD:Senellart}, but appeared after sample accidental freezing.}

\begin{figure*}[htp]
\begin{center}
\includegraphics[width=.85\linewidth]{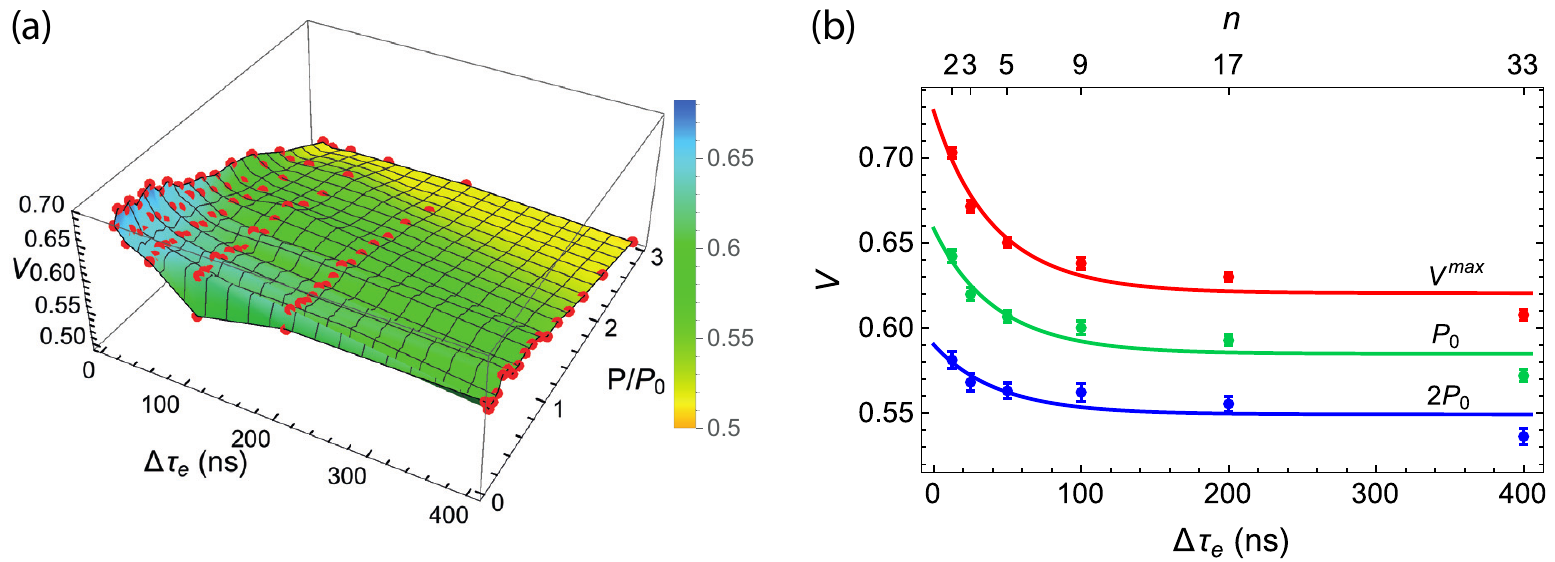}
\caption{Power- and temporal-dependent two-photon interference. a) Over ${>}100$ measured visibilities (red points) showing conclusive quantum interference, i.e. $V{>}0.5$, at all measured powers and timescales. Coloured surface is an interpolation to the data. b) Fitted values of $\overline{V}$ at different $\Delta\tau_e$ (bottom axis), for $P{=}0$ (red), $P{=}P_0$ (green), and $P{=}2P_0$ (blue), showing interference between a first and $n$-th consecutive emitted photon (top axis). Curves are fits to our model in Eq.~(\ref{eq:2}).}
\label{fig:3}
\end{center}
\end{figure*}
We now explore the indistinguishability of photons emitted by \emph{Device~$1$} with various temporal distances. We perform our measurements at $T{=}13$~K to reduce phonon-induced dephasing~\cite{TPIDephasing:Kamp}, which is sufficiently close to the quantum dot cavity resonance at $T{=}15$~K. Note that contrary to most reports, the phonon sideband here is not filtered out by the $0.85$~nm bandpass filter used to further suppress the laser light. Figure~\ref{fig:2}a depicts our experimental setup. Single-photons are injected into an unbalanced Mach-Zehnder interferometer with a variable fibre-based path-length difference designed to match---by using multiple fibres of distinct lengths---an integer multiple of $12.5$~ns up to $400$~ns. Polarisation control---polariser (Pol) and a half-wave plate (HWP)---and a polarising beamsplitter (PBS) behave as a beamsplitter with tuneable reflectivity, thus balancing the photon-flux entering the interference point inside a fibre-beamsplitter (FBS) closing the Mach-Zehnder configuration. Quarter-wave plates (QWPs) and HWPs are used to tune the polarisation of interfering photons in parallel or orthogonal configuration. Time-correlation histograms from the output of this interferometer reveal the indistinguishability of photons emitted with a temporal distance $\Delta\tau_e$. Fully distinguishable photons---e.g., with orthogonal polarisation---meeting at a $50{:}50$ beamsplitter result in a $50$~\% probability of being detected simultaneously at the output of the beamsplitter. This results in the peak around $\Delta t{=}0$ of the time-correlation measurement being about half of those at $\Delta t${\textgreater}$0$, with the exception of peaks at $\Delta t{=}\Delta\tau_e$, which larger suppression indicates that the interfering photons were emitted with a temporal distance $\Delta\tau_e$. In general it can be shown for a pure single-photon source, see Supplementary Material, that the areas $A_{\Delta t}$ centered around $\Delta t$ are given by $A_{k}{=}N$, $A_{-\Delta\tau_e}{=}N(1{-}\mathcal{R}^2)$, $A_{\Delta\tau_e}{=}N(1{-}\mathcal{T}^2)$, and $A_0{=}N\left(\left(\mathcal{R}^2+\mathcal{T}^2\right)-2\mathcal{R}\mathcal{T}V\right)$, where $k{=}\pm12.5\text{~ns},\pm25\text{~ns},...$, and excludes peaks at $\pm\Delta\tau_e$, $N$ is an integration constant{, $\mathcal{R}$ is the beamsplitter reflectivity, and $\mathcal{T}{=}1{-}\mathcal{R}$.}

We use the visibility $V$ to quantify the degree of indistinguishability of the source. Since the measured visibility depends both on the photon source and on the apparatus used to characterise it the latter must be accounted for. Ideally the apparatus is a beamsplitter of reflectivity $\mathcal{R}$=0.5; in our experiment $\mathcal{R}{=}0.471$, {$\mathcal{T}{=}0.529$}, and the visibility $V$ is thus, 
	\begin{equation}\label{eq:1}
		V=\frac{\mathcal{R}^2+\mathcal{T}^2-A_0{/}A}{2\mathcal{R}\mathcal{T}},
	\end{equation}
where $A$ is taken as the average value of $A_k$. Note that since the $g^{(2)}(0)$ values are intrinsic to the source, and hence affect any process in which we wish to use it, we do not correct for non-zero $g^{(2)}(0)$ in Eq.~(\ref{eq:1}). The deduced $V$ therefore corresponds to the raw two-photon interference visibility, and quantifies the degree of photon indistinguishability.

Figure~\ref{fig:2}b shows histograms for the indistinguishability of orthogonal- and parallel-polarised photons at $\Delta\tau_e{=}50$~ns and $P{=}P_0$. In virtue of Eq.~(\ref{eq:1}), and measured $\mathcal{R}{=}0.471$, we obtain $V^{P_0}_{50\text{ns}}{=}(0.71{\pm}0.01)\%$ in orthogonal configuration (red histogram), and $V^{P_0}_{50\text{ns}}{=}(60.31{\pm}0.60)\%$ for parallel-polarised photons (blue histogram), where $V^{P}_{\Delta\tau_e}$ denotes visibility taken at a power $P$ and temporal delay $\Delta\tau_e$. We observe higher visibilities at lower powers and shorter delays. For instance, the measurements in Fig.~\ref{fig:2}c were taken at $P{=}0.5P_0$, and reveal $V^{0.5P_0}_{12.5\text{ns}}{=}(67.52{\pm}0.78)\%$ at a temporal delay (blue histogram) of $\Delta\tau_e{=}12.5$~ns. Remarkably, we find that indistinguishability is robust in the temporal domain. Even after $33$ consecutive emitted photons (orange histogram), at $\Delta\tau_e{=}400$~ns, it only decreases to $V^{0.5P_0}_{400\text{ns}}{=}(59.97{\pm}0.76)\%$. That is, less than $8\%$ visibility decrease in $\sim400$~ns. {All $V$ values with the non-resonant scheme are obtained without any background correction.}

To thoroughly examine the indistinguishability properties of \emph{Device~$1$}, we carried out power- and temporal-dependent measurements, see Fig.~\ref{fig:3}a. All these measured $V$ are within the $50\%{-}70\%$ range, thus showing conclusive quantum interference at all measured powers and timescales. The large available photon flux allows us to gather more than $100$ visibility values with measurement errors sufficiently small to identify an interesting behaviour in this narrow visibility range. At any given $\Delta\tau_e$, $V$ is linear in $P$, see Supplementary Material, and we simply use $\overline{V}{=}V_{\Delta\tau_e}^{max}{+}m_{\Delta\tau_e}P$ to characterise the $P$-dependence of $V$ at fixed $\Delta\tau_e$. Conversely, at fixed $P$, $V$ decreases monotonically and asymptotically in $\Delta\tau_e$, flattening to fixed values at longer timescales.

\begin{figure*}[htp]
\begin{center}
\includegraphics[width=.8\linewidth]{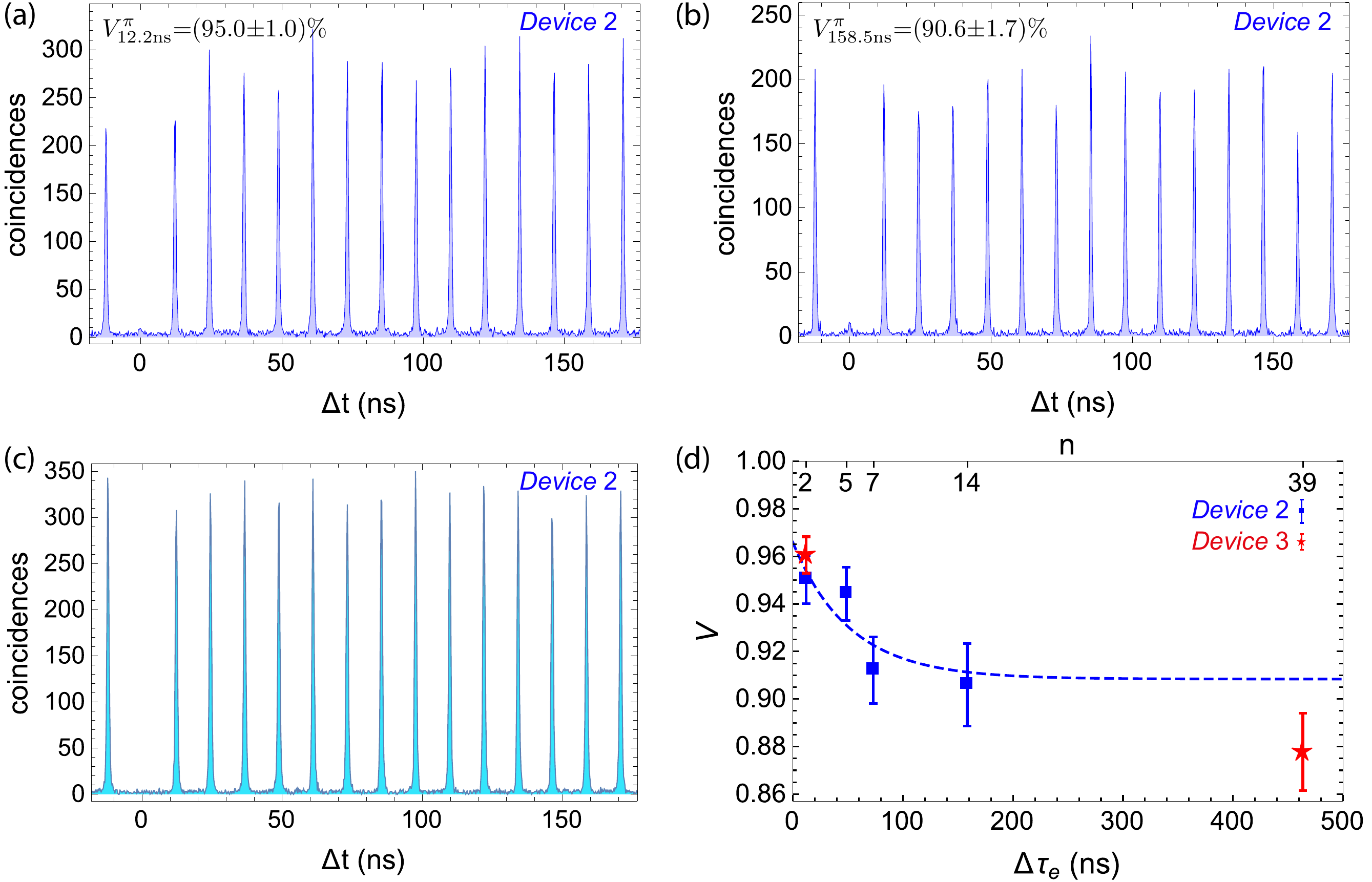}
\caption{Temporal-dependent indistinguishability under strictly resonant excitation. {Two-photon interference histograms with \emph{Device~2} of parallel-polarised photons at a) $\Delta\tau_e{=}12.2$~ns, and b) $\Delta\tau_e{=}158.5$~ns, under a $\pi$-pulse preparation. c) Second-order autocorrelation measurement at $\pi$-pulse. d) Indistinguishability between a first and $n$-th consecutive emitted photon from \emph{Device~2} (blue) and \emph{Device~3} (red). Indistinguishability remains robust in the temporal domain, decreasing only by $4.4\%$ in $\sim159$~ns (down to 90.6\%) for \emph{Device~2}, and by $8.3\%$ in $\sim463$~ns (down to 87.8\%) for \emph{Device~2} . The curve is a fit of the data from \emph{Device~2} to Eq.~(\ref{eq:2}).}}
\label{fig:4}
\end{center}
\end{figure*}
We model this behaviour by considering a time-dependent wandering of the spectral line as the origin of the temporal modulation. That is, the frequency of every emitted photon $\omega(t){=}\omega_0{+}\delta\omega(t)$ varies in time according to some wandering function $\delta\omega(t)$ occurring in timescales much larger than the photon lifetime. Our problem is then equivalent to finding the mutual interference visibility between independent sources with finite frequency detuning~\cite{cavEnhTPI:Giesz}, which is given by $V(0)/\left( 1+ \delta\omega_r^2 \right)$ in the case where $V(0)$ is the degree of indistinguishability for each source alone (equal value for both), and $\delta\omega_r$ is the ratio of the frequency detuning to the spectral linewidth of the sources (equal linewidth for both). If this mismatch arises due to spectral wandering within the same source, then the time-averaged relative detuning squared is given by $2\delta\omega_r^2\left(1{-}\exp\left({{-}\Delta\tau_e/\tau_c}\right)\right)$, with $\tau_c$ a characteristic wandering timescale, see Supplementary Material for more details. We thus derive the visibility of temporally-distant photons:	
	\begin{equation}\label{eq:2}
		V\left(\Delta\tau_e\right)=\frac{V(0)}{1+{2\delta\omega_r}^2\left(1-e^{-\Delta\tau_e/\tau_c}\right)}.
	\end{equation}
To obtain a statistically meaningful temporal behaviour, we used the fitted values of $\overline{V}$ at different $\Delta\tau_e$, for powers $P{=}0$, $P{=}P_0$, and $P{=}2P_0$. These values are plotted in Fig.~\ref{fig:3}b and are in good agreement with our model in Eq.~(\ref{eq:2}). In the limit of low powers, we obtain $V(0){=}(72.8\pm2.4)\%$, $\tau_c{=}(45.5\pm19.1)$~ns, and $\delta\omega_r{=}(29.4\pm3.1)\%$; whereas at high powers, at $P{=}2P_0$, these parameters are $V(0){=}(59.0\pm2.0)\%$, and $\delta\omega_r{=}(19.3\pm4.5)\%$.  {The maximum degree of indistinguishability $V(0)$ decreases only by $13.8\%$ with increasing power, evidencing a slight increase of pure dephasing of the exciton transition. On the contrary, the relative amplitude of the spectral wandering decreases by 34\%, evidencing that spectral diffusion is significantly reduced at higher powers, as recently observed in nanowire based devices~\cite{spectral-diff-nanowire}. Note that the large relative error in $\tau_c$ is due to a small relative decay in $V$, an uncertainty that increases with increasing power}. Thus---although it is reasonable to assume that $\tau_c$ itself is power-dependent---we extracted $\tau_c$ only at $P{=}0$ and used it as a fixed parameter for the fits at higher powers.

{The decrease of the indistinguishability by few percents for temporally distant photons demonstrates a very limited spectral diffusion in our micropillar devices. This observation is in striking contrast to previous measurements on single photon sources based on alternative approaches for efficient photon extraction, such as nanowires~\cite{spectral-diff-nanowire}, or micro lenses~\cite{QD:Thoma}. A significantly lower stability of the electrostatic environment of the QD can reasonably be attributed  to the close proximity of free surfaces in the latter. Indeed, as indicated by the observation of three emission lines from the same QD, even the micropillar devices under study do not provide a fully stable charge state for the QDs, an effect that we observe to be dependent on the quality of the etched surfaces. This makes strictly resonant spectroscopy difficult without an additional non-resonant excitation, a situation also observed in other micropillar devices~\cite{10000:JWP}.}

{Therefore, to explore the indistinguishability of temporally-distant photons under strictly resonant excitation, we turn to electrically controlled micropillars and present data on two devices, \emph{Device~$2$} and \emph{Device~$3$}.  These devices consist of quantum dots deterministically coupled to micropillars embedded in cylindrical gated structures with $p$- and $n$-contacts respectively defined on the top and bottom sides of the device, resulting in an effective $p${-}$i${-}$n$ diode structure onto which an electric field can be applied. (See Ref.~\cite{nearOpt:Senellart} for a detailed description of the device).} We perform our measurements at $T{=}9$~K and tune the emission into cavity-resonance via an applied bias voltage of ${-}0.3$~V. This sample is cooled by gas exchange in a closed-cycle cryostat, and is pumped by shaped $15$~ps laser pulses at $82$~MHz repetition rate. The experimental setup used for photon collection is reported in Ref.~\cite{nearOpt:Senellart}, and the appartus used for the temporal-dependent measurements is conceptually identical to that in Fig.~\ref{fig:2}a.

{Resonant-excitation allows us to probe two-photon interference in a regime excelling in indistinguishability performance. Indeed, for \emph{Device~$2$} we obtain $V^{\pi}_{12.2\text{ns}}{=}(95.0{\pm}1.0)\%$ at a short temporal separation, decreasing only to $V^{\pi}_{158.5\text{ns}}{=}(90.6{\pm}1.7)\%$ at long timescales, see Figs.~\ref{fig:4}a, and \ref{fig:4}b. We observe a high single-photon purity quantified by $g^{(2)}(0){=}0.015\pm0.007$ at $\pi$-pulse, see Fig.~\ref{fig:4}c, where the non-vanishing $g^{(2)}(0)$ primarily consists of background noise and thus a value $1{-}g^{(2)}(0)$ of $98.5\%$ represents a lower bound on the intrinsic single-photon purity. Indistinguishability measurements at various temporal distances, see Fig.~\ref{fig:4}d, reveal plateaus at high values: Up to a first and fourteenth photon, separated by $\sim150$~ns, exhibit an indistinguishability greater than $90\%$. The curve is a fit to Eq.~(\ref{eq:2}), with a maximum indistinguishability value of $V(0){=}96.6\%$, $\tau_c{=}54.4$~ns, and $\delta\omega_r{=}17.8\%$. The reproducibility of our results, thanks to a deterministic fabrication, is evidenced by similar indistinguishability values obtained on \emph{Device~$3$}: $V^{\pi}_{12.2\text{ns}}{=}(96.1\pm0.8)\%$ at a short temporal delay, and $V^{\pi}_{463\text{ns}}{=}(87.8\pm1.6)\%$ for a first and thirty-ninth photon separated by 463~ns. These values of indistinguishability are corrected for the measured background noise arising from detector dark counts: The experimental setup used for these resonant-excitation measurements presents a low collection efficiency, thus an integration of raw detected counts that includes the background noise, which at zero delay is as large as non-vanishing counts due to photon dinstinguishability, would under-estimate the intrinsic degrees of indistinguishability in our devices, see the Supplementary Material for details on this method. No correction for non-vanishing $g^{(2)}(0)$ was included.} 

Note that a high absolute brightness with this recently developed technology is yet to be achieved. {However, since the mode profile of connected pillars is the same as isolated ones \cite{spDetElecCont:Senellart} and a photon extraction efficiency at the first lens of $65\%$ has been reported on this sample~\cite{nearOpt:Senellart}, the same experimental methods as before should allow even higher absolute efficiencies than the $14\%$ reported here.}

{We provided here strong evidence that our sources emit long streams of indistinguishable photons. Under non-resonant excitation, even a first and a thirty-third consecutive photon, separated by $400$~ns, display conclusive quantum interference. For a fixed pump power, photon indistinguishability decreases only a few percent---about $8\%$ at low powers and less than $4\%$ at higher powers---before flattening to fixed values at longer timescales. This contrasts favourably to previous works, where photon indistinguishability has been observed to decrease by $40\%$ in only $10$~ns~\cite{QD:Thoma}. Moreover, under strictly-resonant excitation, photon indistinguishability between a first and thirty-ninth photon remained at $88\%$. Interestingly, the observation of only small reductions in the temporal domain indicate that non-unity indistinguishability under non-resonant excitation is mainly caused by homogenous broadening of the spectral linewidth (governing coherence times at short temporal delays), and a limited inhomogeneous broadening (governing effective coherence times at longer temporal delays). The relative amplitude of the spectral diffusion at saturation is similar for both resonant and non-resonant excitation. However, \emph{Device~$1$} operates in a limited Purcell regime whereas \emph{Devices}~$2$ and $3$ operate with a Purcell factor around $7{-}10$, leading to an increased radiative exciton linewidth. From this, we conclude that, although the application of an electrical bias in n-i-p diode structures allows a good control of the QD charge states, it does not lead to a significant decrease in the spectral wandering phenomena. The excellent indistinguishability observed in \emph{Devices} 2 and 3 arises mainly from reduced pure dephasing of the exciton state, increased Purcell factor and reduced time jitter in a resonant excitation scheme.}

{Our reported indistinguishability values correspond to the longest temporal delays here studied, at a particular pump repetition rate of $80$~MHz:} It only represents a lower bound on the number of photons we can generate---limited by radiative lifetimes in the order of a few hundred picoseconds---that can be further used in quantum information processing protocols with solid-state sources~\cite{entGate:Senellart}. Previous works investigating noise spectra in resonance fluorescence have shown evidence of long streams of near transform-limited photons~\cite{indNoise:Warburton} in timescales potentially reaching seconds~\cite{transLimitedQD:Warburton}. {In fact, \emph{Device~$2$} has recently been shown to emit photons with near transform-limited linewidth in a millisecond timescale~\cite{fewPhoton:Giesz}, in which case we would expect that our devices are producing at least hundreds of thousands of highly indistinguishable single-photons.}

Our findings are especially relevant in implementations with time-bin encoded degrees of freedom, such as some recently proposed schemes of linear-optics quantum computing with time-bin encoding~\cite{LOQCSPM:Walmsley,LOQCTLoop:Rohde}, where the indistinguishability of temporally-distant photons will directly determine quantum fidelities of the implemented protocols. Scaling solid-state multi-photon sources by combining multiple independent emitters remains challenging, as atomic growth accuracy or complex individual electric control over multiple devices is needed. These requirements can be circumvented by making use of a single photon source emitting a long temporal stream of highly indistinguishable photons that can be demultiplexed by fast active optics.

{A high absolute brightness will be critical for successfully implementing multi-photon experiments with these sources, where their downconversion counterparts currently require experimental runs of hundreds of hours~\cite{ScattBS:Sciarrino,6photon:Guo}. The key role of high emission yields in these devices has been made explicit in the recent demonstration of a solid-state based multi-photon experiment~\cite{BS:Loredo}, realised with \emph{Device~$1$}, where integration times outperformed those in equivalent downconversion implementations by two-orders of magnitude. Achieving high absolute efficiencies, and thus allowing the scaling of multi-photon experiments to larger photon numbers, becomes feasible due to Purcell-enhancement of \emph{deterministically}-coupled quantum dot-micropillar devices~\cite{indQD:Senellart,spDetElecCont:Senellart,cavEnhTPI:Giesz,nearOpt:Senellart,detPillar75:Hofling}}. {This} necessary condition {is} unlikely to be found by chance with non-deterministic approaches, with reported~\cite{detPillar75:Hofling} device yields of $\sim$0.01\%~\cite{10000:JWP}. Thus, the deterministic fabrication, high absolute brightness, and long timescale indistinguishability of our devices will enable large-scale applications that have been heretofore impossible.

\subsection*{Funding Information}

Centre for Engineered Quantum Systems (CE110001013); Centre for Quantum Computation and Communication Technology (CE110001027); Asian Office of Aerospace Research and Development (FA2386-13-1-4070); ARC Discovery Early Career Research Award (DE120101899); ERC Starting Grant (277885 QD-CQED); French Agence Nationale pour la Recherche (ANR DELIGHT, ANR USSEPP); French RENATECH network Labex NanoSaclay; European UnionÕs Seventh Framework Programme FP7 (618078 WASPS)

\subsection*{Acknowledgments}

J.~C.~L. and A.~G.~W. thank the team from the Austrian Institute of Technology for kindly providing the time-tagging modules. M. P. A. thanks Halina Rubinsztein-Dunlop for the generous loan of equipment. The CNRS-LPN authors are very thankful to Anna Nowak for her help with the technology. A. G. W. acknowledges support from a UQ Vice ChancellorÕs Research and Teaching Fellowship

\bibliography{refV2.bib}

\begin{thebibliography}{48}%
\makeatletter
\providecommand \@ifxundefined [1]{%
 \@ifx{#1\undefined}
}%
\providecommand \@ifnum [1]{%
 \ifnum #1\expandafter \@firstoftwo
 \else \expandafter \@secondoftwo
 \fi
}%
\providecommand \@ifx [1]{%
 \ifx #1\expandafter \@firstoftwo
 \else \expandafter \@secondoftwo
 \fi
}%
\providecommand \natexlab [1]{#1}%
\providecommand \enquote  [1]{``#1''}%
\providecommand \bibnamefont  [1]{#1}%
\providecommand \bibfnamefont [1]{#1}%
\providecommand \citenamefont [1]{#1}%
\providecommand \href@noop [0]{\@secondoftwo}%
\providecommand \href [0]{\begingroup \@sanitize@url \@href}%
\providecommand \@href[1]{\@@startlink{#1}\@@href}%
\providecommand \@@href[1]{\endgroup#1\@@endlink}%
\providecommand \@sanitize@url [0]{\catcode `\\12\catcode `\$12\catcode
  `\&12\catcode `\#12\catcode `\^12\catcode `\_12\catcode `\%12\relax}%
\providecommand \@@startlink[1]{}%
\providecommand \@@endlink[0]{}%
\providecommand \url  [0]{\begingroup\@sanitize@url \@url }%
\providecommand \@url [1]{\endgroup\@href {#1}{\urlprefix }}%
\providecommand \urlprefix  [0]{URL }%
\providecommand \Eprint [0]{\href }%
\providecommand \doibase [0]{http://dx.doi.org/}%
\providecommand \selectlanguage [0]{\@gobble}%
\providecommand \bibinfo  [0]{\@secondoftwo}%
\providecommand \bibfield  [0]{\@secondoftwo}%
\providecommand \translation [1]{[#1]}%
\providecommand \BibitemOpen [0]{}%
\providecommand \bibitemStop [0]{}%
\providecommand \bibitemNoStop [0]{.\EOS\space}%
\providecommand \EOS [0]{\spacefactor3000\relax}%
\providecommand \BibitemShut  [1]{\csname bibitem#1\endcsname}%
\let\auto@bib@innerbib\@empty
\bibitem [{\citenamefont {Hong}\ \emph {et~al.}(1987)\citenamefont {Hong},
  \citenamefont {Ou},\ and\ \citenamefont {Mandel}}]{HOM:Mandel}%
  \BibitemOpen
  \bibfield  {author} {\bibinfo {author} {\bibfnamefont {C.~K.}\ \bibnamefont
  {Hong}}, \bibinfo {author} {\bibfnamefont {Z.~Y.}\ \bibnamefont {Ou}}, \ and\
  \bibinfo {author} {\bibfnamefont {L.}~\bibnamefont {Mandel}},\ }\href
  {\doibase 10.1103/PhysRevLett.59.2044} {\bibfield  {journal} {\bibinfo
  {journal} {Phys. Rev. Lett.}\ }\textbf {\bibinfo {volume} {59}},\ \bibinfo
  {pages} {2044} (\bibinfo {year} {1987})}\BibitemShut {NoStop}%
\bibitem [{\citenamefont {Legero}\ \emph {et~al.}(2004)\citenamefont {Legero},
  \citenamefont {Wilk}, \citenamefont {Hennrich}, \citenamefont {Rempe},\ and\
  \citenamefont {Kuhn}}]{indAtom:Kuhn}%
  \BibitemOpen
  \bibfield  {author} {\bibinfo {author} {\bibfnamefont {T.}~\bibnamefont
  {Legero}}, \bibinfo {author} {\bibfnamefont {T.}~\bibnamefont {Wilk}},
  \bibinfo {author} {\bibfnamefont {M.}~\bibnamefont {Hennrich}}, \bibinfo
  {author} {\bibfnamefont {G.}~\bibnamefont {Rempe}}, \ and\ \bibinfo {author}
  {\bibfnamefont {A.}~\bibnamefont {Kuhn}},\ }\href {\doibase
  10.1103/PhysRevLett.93.070503} {\bibfield  {journal} {\bibinfo  {journal}
  {Phys. Rev. Lett.}\ }\textbf {\bibinfo {volume} {93}},\ \bibinfo {pages}
  {070503} (\bibinfo {year} {2004})}\BibitemShut {NoStop}%
\bibitem [{\citenamefont {Duan}\ and\ \citenamefont
  {Monroe}(2010)}]{indIonR:Monroe}%
  \BibitemOpen
  \bibfield  {author} {\bibinfo {author} {\bibfnamefont {L.-M.}\ \bibnamefont
  {Duan}}\ and\ \bibinfo {author} {\bibfnamefont {C.}~\bibnamefont {Monroe}},\
  }\href {\doibase 10.1103/RevModPhys.82.1209} {\bibfield  {journal} {\bibinfo
  {journal} {Rev. Mod. Phys.}\ }\textbf {\bibinfo {volume} {82}},\ \bibinfo
  {pages} {1209} (\bibinfo {year} {2010})}\BibitemShut {NoStop}%
\bibitem [{\citenamefont {Bernien}\ \emph {et~al.}(2012)\citenamefont
  {Bernien}, \citenamefont {Childress}, \citenamefont {Robledo}, \citenamefont
  {Markham}, \citenamefont {Twitchen},\ and\ \citenamefont
  {Hanson}}]{indNV:Hanson}%
  \BibitemOpen
  \bibfield  {author} {\bibinfo {author} {\bibfnamefont {H.}~\bibnamefont
  {Bernien}}, \bibinfo {author} {\bibfnamefont {L.}~\bibnamefont {Childress}},
  \bibinfo {author} {\bibfnamefont {L.}~\bibnamefont {Robledo}}, \bibinfo
  {author} {\bibfnamefont {M.}~\bibnamefont {Markham}}, \bibinfo {author}
  {\bibfnamefont {D.}~\bibnamefont {Twitchen}}, \ and\ \bibinfo {author}
  {\bibfnamefont {R.}~\bibnamefont {Hanson}},\ }\href {\doibase
  10.1103/PhysRevLett.108.043604} {\bibfield  {journal} {\bibinfo  {journal}
  {Phys. Rev. Lett.}\ }\textbf {\bibinfo {volume} {108}},\ \bibinfo {pages}
  {043604} (\bibinfo {year} {2012})}\BibitemShut {NoStop}%
\bibitem [{\citenamefont {Kiraz}\ \emph {et~al.}(2005)\citenamefont {Kiraz},
  \citenamefont {Ehrl}, \citenamefont {Hellerer}, \citenamefont
  {M\"ustecapl\ifmmode \imath \else \i \fi{}o\ifmmode~\breve{g}\else
  \u{g}\fi{}lu}, \citenamefont {Br\"auchle},\ and\ \citenamefont
  {Zumbusch}}]{indMol:Zumbusch}%
  \BibitemOpen
  \bibfield  {author} {\bibinfo {author} {\bibfnamefont {A.}~\bibnamefont
  {Kiraz}}, \bibinfo {author} {\bibfnamefont {M.}~\bibnamefont {Ehrl}},
  \bibinfo {author} {\bibfnamefont {T.}~\bibnamefont {Hellerer}}, \bibinfo
  {author} {\bibfnamefont {O.~E.}\ \bibnamefont {M\"ustecapl\ifmmode \imath
  \else \i \fi{}o\ifmmode~\breve{g}\else \u{g}\fi{}lu}}, \bibinfo {author}
  {\bibfnamefont {C.}~\bibnamefont {Br\"auchle}}, \ and\ \bibinfo {author}
  {\bibfnamefont {A.}~\bibnamefont {Zumbusch}},\ }\href {\doibase
  10.1103/PhysRevLett.94.223602} {\bibfield  {journal} {\bibinfo  {journal}
  {Phys. Rev. Lett.}\ }\textbf {\bibinfo {volume} {94}},\ \bibinfo {pages}
  {223602} (\bibinfo {year} {2005})}\BibitemShut {NoStop}%
\bibitem [{\citenamefont {Lang}\ \emph {et~al.}(2013)\citenamefont {Lang},
  \citenamefont {Eichler}, \citenamefont {Steffen}, \citenamefont {Fink},
  \citenamefont {Woolley}, \citenamefont {Blais},\ and\ \citenamefont
  {Wallraff}}]{indSCirc:Wallraff}%
  \BibitemOpen
  \bibfield  {author} {\bibinfo {author} {\bibfnamefont {C.}~\bibnamefont
  {Lang}}, \bibinfo {author} {\bibfnamefont {C.}~\bibnamefont {Eichler}},
  \bibinfo {author} {\bibfnamefont {L.}~\bibnamefont {Steffen}}, \bibinfo
  {author} {\bibfnamefont {J.~M.}\ \bibnamefont {Fink}}, \bibinfo {author}
  {\bibfnamefont {M.~J.}\ \bibnamefont {Woolley}}, \bibinfo {author}
  {\bibfnamefont {A.}~\bibnamefont {Blais}}, \ and\ \bibinfo {author}
  {\bibfnamefont {A.}~\bibnamefont {Wallraff}},\ }\href
  {http://dx.doi.org/10.1038/nphys2612} {\bibfield  {journal} {\bibinfo
  {journal} {Nat Phys}\ }\textbf {\bibinfo {volume} {9}},\ \bibinfo {pages}
  {345} (\bibinfo {year} {2013})}\BibitemShut {NoStop}%
\bibitem [{\citenamefont {Pittman}\ \emph {et~al.}(2005)\citenamefont
  {Pittman}, \citenamefont {Jacobs},\ and\ \citenamefont
  {Franson}}]{indSPDC:Franson}%
  \BibitemOpen
  \bibfield  {author} {\bibinfo {author} {\bibfnamefont {T.}~\bibnamefont
  {Pittman}}, \bibinfo {author} {\bibfnamefont {B.}~\bibnamefont {Jacobs}}, \
  and\ \bibinfo {author} {\bibfnamefont {J.}~\bibnamefont {Franson}},\ }\href
  {\doibase http://dx.doi.org/10.1016/j.optcom.2004.11.027} {\bibfield
  {journal} {\bibinfo  {journal} {Optics Communications}\ }\textbf {\bibinfo
  {volume} {246}},\ \bibinfo {pages} {545 } (\bibinfo {year}
  {2005})}\BibitemShut {NoStop}%
\bibitem [{\citenamefont {Santori}\ \emph {et~al.}(2002)\citenamefont
  {Santori}, \citenamefont {Fattal}, \citenamefont {Vuckovic}, \citenamefont
  {Solomon},\ and\ \citenamefont {Yamamoto}}]{indQD:Yamamoto}%
  \BibitemOpen
  \bibfield  {author} {\bibinfo {author} {\bibfnamefont {C.}~\bibnamefont
  {Santori}}, \bibinfo {author} {\bibfnamefont {D.}~\bibnamefont {Fattal}},
  \bibinfo {author} {\bibfnamefont {J.}~\bibnamefont {Vuckovic}}, \bibinfo
  {author} {\bibfnamefont {G.~S.}\ \bibnamefont {Solomon}}, \ and\ \bibinfo
  {author} {\bibfnamefont {Y.}~\bibnamefont {Yamamoto}},\ }\href
  {http://dx.doi.org/10.1038/nature01086} {\bibfield  {journal} {\bibinfo
  {journal} {Nature}\ }\textbf {\bibinfo {volume} {419}},\ \bibinfo {pages}
  {594} (\bibinfo {year} {2002})}\BibitemShut {NoStop}%
\bibitem [{\citenamefont {Gazzano}\ \emph
  {et~al.}(2013{\natexlab{a}})\citenamefont {Gazzano}, \citenamefont
  {Michaelis~de Vasconcellos}, \citenamefont {Arnold}, \citenamefont {Nowak},
  \citenamefont {Galopin}, \citenamefont {Sagnes}, \citenamefont {Lanco},
  \citenamefont {Lema{\^\i}tre},\ and\ \citenamefont
  {Senellart}}]{indQD:Senellart}%
  \BibitemOpen
  \bibfield  {author} {\bibinfo {author} {\bibfnamefont {O.}~\bibnamefont
  {Gazzano}}, \bibinfo {author} {\bibfnamefont {S.}~\bibnamefont {Michaelis~de
  Vasconcellos}}, \bibinfo {author} {\bibfnamefont {C.}~\bibnamefont {Arnold}},
  \bibinfo {author} {\bibfnamefont {A.}~\bibnamefont {Nowak}}, \bibinfo
  {author} {\bibfnamefont {E.}~\bibnamefont {Galopin}}, \bibinfo {author}
  {\bibfnamefont {I.}~\bibnamefont {Sagnes}}, \bibinfo {author} {\bibfnamefont
  {L.}~\bibnamefont {Lanco}}, \bibinfo {author} {\bibfnamefont
  {A.}~\bibnamefont {Lema{\^\i}tre}}, \ and\ \bibinfo {author} {\bibfnamefont
  {P.}~\bibnamefont {Senellart}},\ }\href
  {http://dx.doi.org/10.1038/ncomms2434} {\bibfield  {journal} {\bibinfo
  {journal} {Nat Commun}\ }\textbf {\bibinfo {volume} {4}},\ \bibinfo {pages}
  {1425} (\bibinfo {year} {2013}{\natexlab{a}})}\BibitemShut {NoStop}%
\bibitem [{\citenamefont {O'Brien}\ \emph {et~al.}(2009)\citenamefont
  {O'Brien}, \citenamefont {Furusawa},\ and\ \citenamefont
  {Vuckovic}}]{QuantTech:Obrien}%
  \BibitemOpen
  \bibfield  {author} {\bibinfo {author} {\bibfnamefont {J.~L.}\ \bibnamefont
  {O'Brien}}, \bibinfo {author} {\bibfnamefont {A.}~\bibnamefont {Furusawa}}, \
  and\ \bibinfo {author} {\bibfnamefont {J.}~\bibnamefont {Vuckovic}},\ }\href
  {http://dx.doi.org/10.1038/nphoton.2009.229} {\bibfield  {journal} {\bibinfo
  {journal} {Nat. Photon.}\ }\textbf {\bibinfo {volume} {3}},\ \bibinfo {pages}
  {687} (\bibinfo {year} {2009})}\BibitemShut {NoStop}%
\bibitem [{\citenamefont {Bouwmeester}\ \emph {et~al.}(1997)\citenamefont
  {Bouwmeester}, \citenamefont {Pan}, \citenamefont {Mattle}, \citenamefont
  {Eibl}, \citenamefont {Weinfurter},\ and\ \citenamefont
  {Zeilinger}}]{qtel:Zeilinger}%
  \BibitemOpen
  \bibfield  {author} {\bibinfo {author} {\bibfnamefont {D.}~\bibnamefont
  {Bouwmeester}}, \bibinfo {author} {\bibfnamefont {J.-W.}\ \bibnamefont
  {Pan}}, \bibinfo {author} {\bibfnamefont {K.}~\bibnamefont {Mattle}},
  \bibinfo {author} {\bibfnamefont {M.}~\bibnamefont {Eibl}}, \bibinfo {author}
  {\bibfnamefont {H.}~\bibnamefont {Weinfurter}}, \ and\ \bibinfo {author}
  {\bibfnamefont {A.}~\bibnamefont {Zeilinger}},\ }\href
  {http://dx.doi.org/10.1038/37539} {\bibfield  {journal} {\bibinfo  {journal}
  {Nature}\ }\textbf {\bibinfo {volume} {390}},\ \bibinfo {pages} {575}
  (\bibinfo {year} {1997})}\BibitemShut {NoStop}%
\bibitem [{\citenamefont {Wang}\ \emph {et~al.}(2015)\citenamefont {Wang},
  \citenamefont {Cai}, \citenamefont {Su}, \citenamefont {Chen}, \citenamefont
  {Wu}, \citenamefont {Li}, \citenamefont {Liu}, \citenamefont {Lu},\ and\
  \citenamefont {Pan}}]{qtelM:Pan}%
  \BibitemOpen
  \bibfield  {author} {\bibinfo {author} {\bibfnamefont {X.-L.}\ \bibnamefont
  {Wang}}, \bibinfo {author} {\bibfnamefont {X.-D.}\ \bibnamefont {Cai}},
  \bibinfo {author} {\bibfnamefont {Z.-E.}\ \bibnamefont {Su}}, \bibinfo
  {author} {\bibfnamefont {M.-C.}\ \bibnamefont {Chen}}, \bibinfo {author}
  {\bibfnamefont {D.}~\bibnamefont {Wu}}, \bibinfo {author} {\bibfnamefont
  {L.}~\bibnamefont {Li}}, \bibinfo {author} {\bibfnamefont {N.-L.}\
  \bibnamefont {Liu}}, \bibinfo {author} {\bibfnamefont {C.-Y.}\ \bibnamefont
  {Lu}}, \ and\ \bibinfo {author} {\bibfnamefont {J.-W.}\ \bibnamefont {Pan}},\
  }\href {http://dx.doi.org/10.1038/nature14246} {\bibfield  {journal}
  {\bibinfo  {journal} {Nature}\ }\textbf {\bibinfo {volume} {518}},\ \bibinfo
  {pages} {516} (\bibinfo {year} {2015})}\BibitemShut {NoStop}%
\bibitem [{\citenamefont {Kwiat}\ \emph {et~al.}(1999)\citenamefont {Kwiat},
  \citenamefont {Waks}, \citenamefont {White}, \citenamefont {Appelbaum},\ and\
  \citenamefont {Eberhard}}]{entSource:Kwiat}%
  \BibitemOpen
  \bibfield  {author} {\bibinfo {author} {\bibfnamefont {P.~G.}\ \bibnamefont
  {Kwiat}}, \bibinfo {author} {\bibfnamefont {E.}~\bibnamefont {Waks}},
  \bibinfo {author} {\bibfnamefont {A.~G.}\ \bibnamefont {White}}, \bibinfo
  {author} {\bibfnamefont {I.}~\bibnamefont {Appelbaum}}, \ and\ \bibinfo
  {author} {\bibfnamefont {P.~H.}\ \bibnamefont {Eberhard}},\ }\href {\doibase
  10.1103/PhysRevA.60.R773} {\bibfield  {journal} {\bibinfo  {journal} {Phys.
  Rev. A}\ }\textbf {\bibinfo {volume} {60}},\ \bibinfo {pages} {R773}
  (\bibinfo {year} {1999})}\BibitemShut {NoStop}%
\bibitem [{\citenamefont {Kim}\ \emph {et~al.}(2006)\citenamefont {Kim},
  \citenamefont {Fiorentino},\ and\ \citenamefont {Wong}}]{entSource:Wong}%
  \BibitemOpen
  \bibfield  {author} {\bibinfo {author} {\bibfnamefont {T.}~\bibnamefont
  {Kim}}, \bibinfo {author} {\bibfnamefont {M.}~\bibnamefont {Fiorentino}}, \
  and\ \bibinfo {author} {\bibfnamefont {F.~N.~C.}\ \bibnamefont {Wong}},\
  }\href {\doibase 10.1103/PhysRevA.73.012316} {\bibfield  {journal} {\bibinfo
  {journal} {Phys. Rev. A}\ }\textbf {\bibinfo {volume} {73}},\ \bibinfo
  {pages} {012316} (\bibinfo {year} {2006})}\BibitemShut {NoStop}%
\bibitem [{\citenamefont {Dousse}\ \emph {et~al.}(2010)\citenamefont {Dousse},
  \citenamefont {Suffczynski}, \citenamefont {Beveratos}, \citenamefont
  {Krebs}, \citenamefont {Lemaitre}, \citenamefont {Sagnes}, \citenamefont
  {Bloch}, \citenamefont {Voisin},\ and\ \citenamefont
  {Senellart}}]{entSource:Senellart}%
  \BibitemOpen
  \bibfield  {author} {\bibinfo {author} {\bibfnamefont {A.}~\bibnamefont
  {Dousse}}, \bibinfo {author} {\bibfnamefont {J.}~\bibnamefont {Suffczynski}},
  \bibinfo {author} {\bibfnamefont {A.}~\bibnamefont {Beveratos}}, \bibinfo
  {author} {\bibfnamefont {O.}~\bibnamefont {Krebs}}, \bibinfo {author}
  {\bibfnamefont {A.}~\bibnamefont {Lemaitre}}, \bibinfo {author}
  {\bibfnamefont {I.}~\bibnamefont {Sagnes}}, \bibinfo {author} {\bibfnamefont
  {J.}~\bibnamefont {Bloch}}, \bibinfo {author} {\bibfnamefont
  {P.}~\bibnamefont {Voisin}}, \ and\ \bibinfo {author} {\bibfnamefont
  {P.}~\bibnamefont {Senellart}},\ }\href
  {http://dx.doi.org/10.1038/nature09148} {\bibfield  {journal} {\bibinfo
  {journal} {Nature}\ }\textbf {\bibinfo {volume} {466}},\ \bibinfo {pages}
  {217} (\bibinfo {year} {2010})}\BibitemShut {NoStop}%
\bibitem [{\citenamefont {Knill}\ \emph {et~al.}(2001)\citenamefont {Knill},
  \citenamefont {Laflamme},\ and\ \citenamefont {Milburn}}]{KLM:KLM}%
  \BibitemOpen
  \bibfield  {author} {\bibinfo {author} {\bibfnamefont {E.}~\bibnamefont
  {Knill}}, \bibinfo {author} {\bibfnamefont {R.}~\bibnamefont {Laflamme}}, \
  and\ \bibinfo {author} {\bibfnamefont {G.~J.}\ \bibnamefont {Milburn}},\
  }\href {http://dx.doi.org/10.1038/35051009} {\bibfield  {journal} {\bibinfo
  {journal} {Nature}\ }\textbf {\bibinfo {volume} {409}},\ \bibinfo {pages}
  {46} (\bibinfo {year} {2001})}\BibitemShut {NoStop}%
\bibitem [{\citenamefont {Kok}\ \emph {et~al.}(2007)\citenamefont {Kok},
  \citenamefont {Munro}, \citenamefont {Nemoto}, \citenamefont {Ralph},
  \citenamefont {Dowling},\ and\ \citenamefont {Milburn}}]{LOQC:Milburn}%
  \BibitemOpen
  \bibfield  {author} {\bibinfo {author} {\bibfnamefont {P.}~\bibnamefont
  {Kok}}, \bibinfo {author} {\bibfnamefont {W.~J.}\ \bibnamefont {Munro}},
  \bibinfo {author} {\bibfnamefont {K.}~\bibnamefont {Nemoto}}, \bibinfo
  {author} {\bibfnamefont {T.~C.}\ \bibnamefont {Ralph}}, \bibinfo {author}
  {\bibfnamefont {J.~P.}\ \bibnamefont {Dowling}}, \ and\ \bibinfo {author}
  {\bibfnamefont {G.~J.}\ \bibnamefont {Milburn}},\ }\href {\doibase
  10.1103/RevModPhys.79.135} {\bibfield  {journal} {\bibinfo  {journal} {Rev.
  Mod. Phys.}\ }\textbf {\bibinfo {volume} {79}},\ \bibinfo {pages} {135}
  (\bibinfo {year} {2007})}\BibitemShut {NoStop}%
\bibitem [{\citenamefont {Michler}\ \emph {et~al.}(2000)\citenamefont
  {Michler}, \citenamefont {Kiraz}, \citenamefont {Becher}, \citenamefont
  {Schoenfeld}, \citenamefont {Petroff}, \citenamefont {Zhang}, \citenamefont
  {Hu},\ and\ \citenamefont {Imamoglu}}]{spDevice:Imamoglu}%
  \BibitemOpen
  \bibfield  {author} {\bibinfo {author} {\bibfnamefont {P.}~\bibnamefont
  {Michler}}, \bibinfo {author} {\bibfnamefont {A.}~\bibnamefont {Kiraz}},
  \bibinfo {author} {\bibfnamefont {C.}~\bibnamefont {Becher}}, \bibinfo
  {author} {\bibfnamefont {W.~V.}\ \bibnamefont {Schoenfeld}}, \bibinfo
  {author} {\bibfnamefont {P.~M.}\ \bibnamefont {Petroff}}, \bibinfo {author}
  {\bibfnamefont {L.}~\bibnamefont {Zhang}}, \bibinfo {author} {\bibfnamefont
  {E.}~\bibnamefont {Hu}}, \ and\ \bibinfo {author} {\bibfnamefont
  {A.}~\bibnamefont {Imamoglu}},\ }\href {\doibase
  10.1126/science.290.5500.2282} {\bibfield  {journal} {\bibinfo  {journal}
  {Science}\ }\textbf {\bibinfo {volume} {290}},\ \bibinfo {pages} {2282}
  (\bibinfo {year} {2000})}\BibitemShut {NoStop}%
\bibitem [{\citenamefont {Vuckovic}\ \emph {et~al.}(2003)\citenamefont
  {Vuckovic}, \citenamefont {Fattal}, \citenamefont {Santori}, \citenamefont
  {Solomon},\ and\ \citenamefont {Yamamoto}}]{spEnhEmis:Yamamoto}%
  \BibitemOpen
  \bibfield  {author} {\bibinfo {author} {\bibfnamefont {J.}~\bibnamefont
  {Vuckovic}}, \bibinfo {author} {\bibfnamefont {D.}~\bibnamefont {Fattal}},
  \bibinfo {author} {\bibfnamefont {C.}~\bibnamefont {Santori}}, \bibinfo
  {author} {\bibfnamefont {G.~S.}\ \bibnamefont {Solomon}}, \ and\ \bibinfo
  {author} {\bibfnamefont {Y.}~\bibnamefont {Yamamoto}},\ }\href {\doibase
  http://dx.doi.org/10.1063/1.1577828} {\bibfield  {journal} {\bibinfo
  {journal} {Applied Physics Letters}\ }\textbf {\bibinfo {volume} {82}},\
  \bibinfo {pages} {3596} (\bibinfo {year} {2003})}\BibitemShut {NoStop}%
\bibitem [{\citenamefont {Pelton}\ \emph {et~al.}(2002)\citenamefont {Pelton},
  \citenamefont {Santori}, \citenamefont {Vuc\ifmmode \breve{}\else
  \u{}\fi{}kovi\ifmmode~\acute{c}\else \'{c}\fi{}}, \citenamefont {Zhang},
  \citenamefont {Solomon}, \citenamefont {Plant},\ and\ \citenamefont
  {Yamamoto}}]{spQDEff:Yamamoto}%
  \BibitemOpen
  \bibfield  {author} {\bibinfo {author} {\bibfnamefont {M.}~\bibnamefont
  {Pelton}}, \bibinfo {author} {\bibfnamefont {C.}~\bibnamefont {Santori}},
  \bibinfo {author} {\bibfnamefont {J.}~\bibnamefont {Vuc\ifmmode \breve{}\else
  \u{}\fi{}kovi\ifmmode~\acute{c}\else \'{c}\fi{}}}, \bibinfo {author}
  {\bibfnamefont {B.}~\bibnamefont {Zhang}}, \bibinfo {author} {\bibfnamefont
  {G.~S.}\ \bibnamefont {Solomon}}, \bibinfo {author} {\bibfnamefont
  {J.}~\bibnamefont {Plant}}, \ and\ \bibinfo {author} {\bibfnamefont
  {Y.}~\bibnamefont {Yamamoto}},\ }\href {\doibase
  10.1103/PhysRevLett.89.233602} {\bibfield  {journal} {\bibinfo  {journal}
  {Phys. Rev. Lett.}\ }\textbf {\bibinfo {volume} {89}},\ \bibinfo {pages}
  {233602} (\bibinfo {year} {2002})}\BibitemShut {NoStop}%
\bibitem [{\citenamefont {Lund-Hansen}\ \emph {et~al.}(2008)\citenamefont
  {Lund-Hansen}, \citenamefont {Stobbe}, \citenamefont {Julsgaard},
  \citenamefont {Thyrrestrup}, \citenamefont {S\"unner}, \citenamefont {Kamp},
  \citenamefont {Forchel},\ and\ \citenamefont {Lodahl}}]{spQDPhCr:Lodahl}%
  \BibitemOpen
  \bibfield  {author} {\bibinfo {author} {\bibfnamefont {T.}~\bibnamefont
  {Lund-Hansen}}, \bibinfo {author} {\bibfnamefont {S.}~\bibnamefont {Stobbe}},
  \bibinfo {author} {\bibfnamefont {B.}~\bibnamefont {Julsgaard}}, \bibinfo
  {author} {\bibfnamefont {H.}~\bibnamefont {Thyrrestrup}}, \bibinfo {author}
  {\bibfnamefont {T.}~\bibnamefont {S\"unner}}, \bibinfo {author}
  {\bibfnamefont {M.}~\bibnamefont {Kamp}}, \bibinfo {author} {\bibfnamefont
  {A.}~\bibnamefont {Forchel}}, \ and\ \bibinfo {author} {\bibfnamefont
  {P.}~\bibnamefont {Lodahl}},\ }\href {\doibase
  10.1103/PhysRevLett.101.113903} {\bibfield  {journal} {\bibinfo  {journal}
  {Phys. Rev. Lett.}\ }\textbf {\bibinfo {volume} {101}},\ \bibinfo {pages}
  {113903} (\bibinfo {year} {2008})}\BibitemShut {NoStop}%
\bibitem [{\citenamefont {Madsen}\ \emph {et~al.}(2014)\citenamefont {Madsen},
  \citenamefont {Ates}, \citenamefont {Liu}, \citenamefont {Javadi},
  \citenamefont {Albrecht}, \citenamefont {Yeo}, \citenamefont {Stobbe},\ and\
  \citenamefont {Lodahl}}]{spQDEffOut:Lodahl}%
  \BibitemOpen
  \bibfield  {author} {\bibinfo {author} {\bibfnamefont {K.~H.}\ \bibnamefont
  {Madsen}}, \bibinfo {author} {\bibfnamefont {S.}~\bibnamefont {Ates}},
  \bibinfo {author} {\bibfnamefont {J.}~\bibnamefont {Liu}}, \bibinfo {author}
  {\bibfnamefont {A.}~\bibnamefont {Javadi}}, \bibinfo {author} {\bibfnamefont
  {S.~M.}\ \bibnamefont {Albrecht}}, \bibinfo {author} {\bibfnamefont
  {I.}~\bibnamefont {Yeo}}, \bibinfo {author} {\bibfnamefont {S.}~\bibnamefont
  {Stobbe}}, \ and\ \bibinfo {author} {\bibfnamefont {P.}~\bibnamefont
  {Lodahl}},\ }\href {\doibase 10.1103/PhysRevB.90.155303} {\bibfield
  {journal} {\bibinfo  {journal} {Phys. Rev. B}\ }\textbf {\bibinfo {volume}
  {90}},\ \bibinfo {pages} {155303} (\bibinfo {year} {2014})}\BibitemShut
  {NoStop}%
\bibitem [{\citenamefont {G\'erard}\ \emph {et~al.}(1998)\citenamefont
  {G\'erard}, \citenamefont {Sermage}, \citenamefont {Gayral}, \citenamefont
  {Legrand}, \citenamefont {Costard},\ and\ \citenamefont
  {Thierry-Mieg}}]{enhancedEm:JMGerard}%
  \BibitemOpen
  \bibfield  {author} {\bibinfo {author} {\bibfnamefont {J.~M.}\ \bibnamefont
  {G\'erard}}, \bibinfo {author} {\bibfnamefont {B.}~\bibnamefont {Sermage}},
  \bibinfo {author} {\bibfnamefont {B.}~\bibnamefont {Gayral}}, \bibinfo
  {author} {\bibfnamefont {B.}~\bibnamefont {Legrand}}, \bibinfo {author}
  {\bibfnamefont {E.}~\bibnamefont {Costard}}, \ and\ \bibinfo {author}
  {\bibfnamefont {V.}~\bibnamefont {Thierry-Mieg}},\ }\href {\doibase
  10.1103/PhysRevLett.81.1110} {\bibfield  {journal} {\bibinfo  {journal}
  {Phys. Rev. Lett.}\ }\textbf {\bibinfo {volume} {81}},\ \bibinfo {pages}
  {1110} (\bibinfo {year} {1998})}\BibitemShut {NoStop}%
\bibitem [{\citenamefont {He}\ \emph {et~al.}(2013)\citenamefont {He},
  \citenamefont {He}, \citenamefont {Wei}, \citenamefont {Wu}, \citenamefont
  {Atature}, \citenamefont {Schneider}, \citenamefont {Hofling}, \citenamefont
  {Kamp}, \citenamefont {Lu},\ and\ \citenamefont {Pan}}]{indQDUnity:Pan}%
  \BibitemOpen
  \bibfield  {author} {\bibinfo {author} {\bibfnamefont {Y.-M.}\ \bibnamefont
  {He}}, \bibinfo {author} {\bibfnamefont {Y.}~\bibnamefont {He}}, \bibinfo
  {author} {\bibfnamefont {Y.-J.}\ \bibnamefont {Wei}}, \bibinfo {author}
  {\bibfnamefont {D.}~\bibnamefont {Wu}}, \bibinfo {author} {\bibfnamefont
  {M.}~\bibnamefont {Atature}}, \bibinfo {author} {\bibfnamefont
  {C.}~\bibnamefont {Schneider}}, \bibinfo {author} {\bibfnamefont
  {S.}~\bibnamefont {Hofling}}, \bibinfo {author} {\bibfnamefont
  {M.}~\bibnamefont {Kamp}}, \bibinfo {author} {\bibfnamefont {C.-Y.}\
  \bibnamefont {Lu}}, \ and\ \bibinfo {author} {\bibfnamefont {J.-W.}\
  \bibnamefont {Pan}},\ }\href {http://dx.doi.org/10.1038/nnano.2012.262}
  {\bibfield  {journal} {\bibinfo  {journal} {Nat Nano}\ }\textbf {\bibinfo
  {volume} {8}},\ \bibinfo {pages} {213} (\bibinfo {year} {2013})}\BibitemShut
  {NoStop}%
\bibitem [{\citenamefont {Wei}\ \emph {et~al.}(2014)\citenamefont {Wei},
  \citenamefont {He}, \citenamefont {Chen}, \citenamefont {Hu}, \citenamefont
  {He}, \citenamefont {Wu}, \citenamefont {Schneider}, \citenamefont {Kamp},
  \citenamefont {H\"ofling}, \citenamefont {Lu},\ and\ \citenamefont
  {Pan}}]{indQD995:Pan}%
  \BibitemOpen
  \bibfield  {author} {\bibinfo {author} {\bibfnamefont {Y.-J.}\ \bibnamefont
  {Wei}}, \bibinfo {author} {\bibfnamefont {Y.-M.}\ \bibnamefont {He}},
  \bibinfo {author} {\bibfnamefont {M.-C.}\ \bibnamefont {Chen}}, \bibinfo
  {author} {\bibfnamefont {Y.-N.}\ \bibnamefont {Hu}}, \bibinfo {author}
  {\bibfnamefont {Y.}~\bibnamefont {He}}, \bibinfo {author} {\bibfnamefont
  {D.}~\bibnamefont {Wu}}, \bibinfo {author} {\bibfnamefont {C.}~\bibnamefont
  {Schneider}}, \bibinfo {author} {\bibfnamefont {M.}~\bibnamefont {Kamp}},
  \bibinfo {author} {\bibfnamefont {S.}~\bibnamefont {H\"ofling}}, \bibinfo
  {author} {\bibfnamefont {C.-Y.}\ \bibnamefont {Lu}}, \ and\ \bibinfo {author}
  {\bibfnamefont {J.-W.}\ \bibnamefont {Pan}},\ }\href {\doibase
  10.1021/nl503081n} {\bibfield  {journal} {\bibinfo  {journal} {Nano Letters}\
  }\textbf {\bibinfo {volume} {14}},\ \bibinfo {pages} {6515} (\bibinfo {year}
  {2014})},\ \bibinfo {note} {pMID: 25357153}\BibitemShut {NoStop}%
\bibitem [{\citenamefont {Ding}\ \emph {et~al.}(2016)\citenamefont {Ding},
  \citenamefont {He}, \citenamefont {Duan}, \citenamefont {Gregersen},
  \citenamefont {Chen}, \citenamefont {Unsleber}, \citenamefont {Maier},
  \citenamefont {Schneider}, \citenamefont {Kamp}, \citenamefont {H\"ofling},
  \citenamefont {Lu},\ and\ \citenamefont {Pan}}]{10000:JWP}%
  \BibitemOpen
  \bibfield  {author} {\bibinfo {author} {\bibfnamefont {X.}~\bibnamefont
  {Ding}}, \bibinfo {author} {\bibfnamefont {Y.}~\bibnamefont {He}}, \bibinfo
  {author} {\bibfnamefont {Z.-C.}\ \bibnamefont {Duan}}, \bibinfo {author}
  {\bibfnamefont {N.}~\bibnamefont {Gregersen}}, \bibinfo {author}
  {\bibfnamefont {M.-C.}\ \bibnamefont {Chen}}, \bibinfo {author}
  {\bibfnamefont {S.}~\bibnamefont {Unsleber}}, \bibinfo {author}
  {\bibfnamefont {S.}~\bibnamefont {Maier}}, \bibinfo {author} {\bibfnamefont
  {C.}~\bibnamefont {Schneider}}, \bibinfo {author} {\bibfnamefont
  {M.}~\bibnamefont {Kamp}}, \bibinfo {author} {\bibfnamefont {S.}~\bibnamefont
  {H\"ofling}}, \bibinfo {author} {\bibfnamefont {C.-Y.}\ \bibnamefont {Lu}}, \
  and\ \bibinfo {author} {\bibfnamefont {J.-W.}\ \bibnamefont {Pan}},\ }\href
  {\doibase 10.1103/PhysRevLett.116.020401} {\bibfield  {journal} {\bibinfo
  {journal} {Phys. Rev. Lett.}\ }\textbf {\bibinfo {volume} {116}},\ \bibinfo
  {pages} {020401} (\bibinfo {year} {2016})}\BibitemShut {NoStop}%
\bibitem [{\citenamefont {Nowak}\ \emph {et~al.}(2014)\citenamefont {Nowak},
  \citenamefont {Portalupi}, \citenamefont {Giesz}, \citenamefont {Gazzano},
  \citenamefont {Dal~Savio}, \citenamefont {Braun}, \citenamefont {Karrai},
  \citenamefont {Arnold}, \citenamefont {Lanco}, \citenamefont {Sagnes},
  \citenamefont {Lema{\^\i}tre},\ and\ \citenamefont
  {Senellart}}]{spDetElecCont:Senellart}%
  \BibitemOpen
  \bibfield  {author} {\bibinfo {author} {\bibfnamefont {A.~K.}\ \bibnamefont
  {Nowak}}, \bibinfo {author} {\bibfnamefont {S.~L.}\ \bibnamefont
  {Portalupi}}, \bibinfo {author} {\bibfnamefont {V.}~\bibnamefont {Giesz}},
  \bibinfo {author} {\bibfnamefont {O.}~\bibnamefont {Gazzano}}, \bibinfo
  {author} {\bibfnamefont {C.}~\bibnamefont {Dal~Savio}}, \bibinfo {author}
  {\bibfnamefont {P.~F.}\ \bibnamefont {Braun}}, \bibinfo {author}
  {\bibfnamefont {K.}~\bibnamefont {Karrai}}, \bibinfo {author} {\bibfnamefont
  {C.}~\bibnamefont {Arnold}}, \bibinfo {author} {\bibfnamefont
  {L.}~\bibnamefont {Lanco}}, \bibinfo {author} {\bibfnamefont
  {I.}~\bibnamefont {Sagnes}}, \bibinfo {author} {\bibfnamefont
  {A.}~\bibnamefont {Lema{\^\i}tre}}, \ and\ \bibinfo {author} {\bibfnamefont
  {P.}~\bibnamefont {Senellart}},\ }\href
  {http://dx.doi.org/10.1038/ncomms4240} {\bibfield  {journal} {\bibinfo
  {journal} {Nat Commun}\ }\textbf {\bibinfo {volume} {5}},\ \bibinfo {pages}
  {3240} (\bibinfo {year} {2014})}\BibitemShut {NoStop}%
\bibitem [{\citenamefont {Somaschi}\ \emph {et~al.}(2016)\citenamefont
  {Somaschi}, \citenamefont {Giesz}, \citenamefont {De~Santis}, \citenamefont
  {Loredo}, \citenamefont {Almeida}, \citenamefont {Hornecker}, \citenamefont
  {Portalupi}, \citenamefont {Grange}, \citenamefont {Ant{\'o}n}, \citenamefont
  {Demory}, \citenamefont {G{\'o}mez}, \citenamefont {Sagnes}, \citenamefont
  {Lanzillotti-Kimura}, \citenamefont {Lema{\'\i}tre}, \citenamefont
  {Auffeves}, \citenamefont {White}, \citenamefont {Lanco},\ and\ \citenamefont
  {Senellart}}]{nearOpt:Senellart}%
  \BibitemOpen
  \bibfield  {author} {\bibinfo {author} {\bibfnamefont {N.}~\bibnamefont
  {Somaschi}}, \bibinfo {author} {\bibfnamefont {V.}~\bibnamefont {Giesz}},
  \bibinfo {author} {\bibfnamefont {L.}~\bibnamefont {De~Santis}}, \bibinfo
  {author} {\bibfnamefont {J.~C.}\ \bibnamefont {Loredo}}, \bibinfo {author}
  {\bibfnamefont {M.~P.}\ \bibnamefont {Almeida}}, \bibinfo {author}
  {\bibfnamefont {G.}~\bibnamefont {Hornecker}}, \bibinfo {author}
  {\bibfnamefont {S.~L.}\ \bibnamefont {Portalupi}}, \bibinfo {author}
  {\bibfnamefont {T.}~\bibnamefont {Grange}}, \bibinfo {author} {\bibfnamefont
  {C.}~\bibnamefont {Ant{\'o}n}}, \bibinfo {author} {\bibfnamefont
  {J.}~\bibnamefont {Demory}}, \bibinfo {author} {\bibfnamefont
  {C.}~\bibnamefont {G{\'o}mez}}, \bibinfo {author} {\bibfnamefont
  {I.}~\bibnamefont {Sagnes}}, \bibinfo {author} {\bibfnamefont {N.~D.}\
  \bibnamefont {Lanzillotti-Kimura}}, \bibinfo {author} {\bibfnamefont
  {A.}~\bibnamefont {Lema{\'\i}tre}}, \bibinfo {author} {\bibfnamefont
  {A.}~\bibnamefont {Auffeves}}, \bibinfo {author} {\bibfnamefont {A.~G.}\
  \bibnamefont {White}}, \bibinfo {author} {\bibfnamefont {L.}~\bibnamefont
  {Lanco}}, \ and\ \bibinfo {author} {\bibfnamefont {P.}~\bibnamefont
  {Senellart}},\ }\href {http://dx.doi.org/10.1038/nphoton.2016.23} {\bibfield
  {journal} {\bibinfo  {journal} {Nat Photon}\ }\textbf {\bibinfo {volume}
  {advance online publication}},\  (\bibinfo {year} {2016})}\BibitemShut
  {NoStop}%
\bibitem [{\citenamefont {Thoma}\ \emph {et~al.}(2016)\citenamefont {Thoma},
  \citenamefont {Schnauber}, \citenamefont {Gschrey}, \citenamefont {Seifried},
  \citenamefont {Wolters}, \citenamefont {Schulze}, \citenamefont
  {Strittmatter}, \citenamefont {Rodt}, \citenamefont {Carmele}, \citenamefont
  {Knorr}, \citenamefont {Heindel},\ and\ \citenamefont
  {Reitzenstein}}]{QD:Thoma}%
  \BibitemOpen
  \bibfield  {author} {\bibinfo {author} {\bibfnamefont {A.}~\bibnamefont
  {Thoma}}, \bibinfo {author} {\bibfnamefont {P.}~\bibnamefont {Schnauber}},
  \bibinfo {author} {\bibfnamefont {M.}~\bibnamefont {Gschrey}}, \bibinfo
  {author} {\bibfnamefont {M.}~\bibnamefont {Seifried}}, \bibinfo {author}
  {\bibfnamefont {J.}~\bibnamefont {Wolters}}, \bibinfo {author} {\bibfnamefont
  {J.-H.}\ \bibnamefont {Schulze}}, \bibinfo {author} {\bibfnamefont
  {A.}~\bibnamefont {Strittmatter}}, \bibinfo {author} {\bibfnamefont
  {S.}~\bibnamefont {Rodt}}, \bibinfo {author} {\bibfnamefont {A.}~\bibnamefont
  {Carmele}}, \bibinfo {author} {\bibfnamefont {A.}~\bibnamefont {Knorr}},
  \bibinfo {author} {\bibfnamefont {T.}~\bibnamefont {Heindel}}, \ and\
  \bibinfo {author} {\bibfnamefont {S.}~\bibnamefont {Reitzenstein}},\ }\href
  {\doibase 10.1103/PhysRevLett.116.033601} {\bibfield  {journal} {\bibinfo
  {journal} {Phys. Rev. Lett.}\ }\textbf {\bibinfo {volume} {116}},\ \bibinfo
  {pages} {033601} (\bibinfo {year} {2016})}\BibitemShut {NoStop}%
\bibitem [{\citenamefont {Dousse}\ \emph {et~al.}(2008)\citenamefont {Dousse},
  \citenamefont {Lanco}, \citenamefont {Suffczy\ifmmode~\acute{n}\else
  \'{n}\fi{}ski}, \citenamefont {Semenova}, \citenamefont {Miard},
  \citenamefont {Lema\^{i}tre}, \citenamefont {Sagnes}, \citenamefont {Roblin},
  \citenamefont {Bloch},\ and\ \citenamefont
  {Senellart}}]{inSituLith:Senellart}%
  \BibitemOpen
  \bibfield  {author} {\bibinfo {author} {\bibfnamefont {A.}~\bibnamefont
  {Dousse}}, \bibinfo {author} {\bibfnamefont {L.}~\bibnamefont {Lanco}},
  \bibinfo {author} {\bibfnamefont {J.}~\bibnamefont
  {Suffczy\ifmmode~\acute{n}\else \'{n}\fi{}ski}}, \bibinfo {author}
  {\bibfnamefont {E.}~\bibnamefont {Semenova}}, \bibinfo {author}
  {\bibfnamefont {A.}~\bibnamefont {Miard}}, \bibinfo {author} {\bibfnamefont
  {A.}~\bibnamefont {Lema\^{i}tre}}, \bibinfo {author} {\bibfnamefont
  {I.}~\bibnamefont {Sagnes}}, \bibinfo {author} {\bibfnamefont
  {C.}~\bibnamefont {Roblin}}, \bibinfo {author} {\bibfnamefont
  {J.}~\bibnamefont {Bloch}}, \ and\ \bibinfo {author} {\bibfnamefont
  {P.}~\bibnamefont {Senellart}},\ }\href {\doibase
  10.1103/PhysRevLett.101.267404} {\bibfield  {journal} {\bibinfo  {journal}
  {Phys. Rev. Lett.}\ }\textbf {\bibinfo {volume} {101}},\ \bibinfo {pages}
  {267404} (\bibinfo {year} {2008})}\BibitemShut {NoStop}%
\bibitem [{\citenamefont {Grilli}\ \emph {et~al.}(1992)\citenamefont {Grilli},
  \citenamefont {Guzzi}, \citenamefont {Zamboni},\ and\ \citenamefont
  {Pavesi}}]{Tscan:Pavesi}%
  \BibitemOpen
  \bibfield  {author} {\bibinfo {author} {\bibfnamefont {E.}~\bibnamefont
  {Grilli}}, \bibinfo {author} {\bibfnamefont {M.}~\bibnamefont {Guzzi}},
  \bibinfo {author} {\bibfnamefont {R.}~\bibnamefont {Zamboni}}, \ and\
  \bibinfo {author} {\bibfnamefont {L.}~\bibnamefont {Pavesi}},\ }\href
  {\doibase 10.1103/PhysRevB.45.1638} {\bibfield  {journal} {\bibinfo
  {journal} {Phys. Rev. B}\ }\textbf {\bibinfo {volume} {45}},\ \bibinfo
  {pages} {1638} (\bibinfo {year} {1992})}\BibitemShut {NoStop}%
\bibitem [{\citenamefont {Hadfield}(2009)}]{detEff:Hadfield}%
  \BibitemOpen
  \bibfield  {author} {\bibinfo {author} {\bibfnamefont {R.~H.}\ \bibnamefont
  {Hadfield}},\ }\href {http://dx.doi.org/10.1038/nphoton.2009.230} {\bibfield
  {journal} {\bibinfo  {journal} {Nat Photon}\ }\textbf {\bibinfo {volume}
  {3}},\ \bibinfo {pages} {696} (\bibinfo {year} {2009})}\BibitemShut {NoStop}%
\bibitem [{\citenamefont {Claudon}\ \emph {et~al.}(2010)\citenamefont
  {Claudon}, \citenamefont {Bleuse}, \citenamefont {Malik}, \citenamefont
  {Bazin}, \citenamefont {Jaffrennou}, \citenamefont {Gregersen}, \citenamefont
  {Sauvan}, \citenamefont {Lalanne},\ and\ \citenamefont
  {Gerard}}]{spQDHighEff:JMGerard}%
  \BibitemOpen
  \bibfield  {author} {\bibinfo {author} {\bibfnamefont {J.}~\bibnamefont
  {Claudon}}, \bibinfo {author} {\bibfnamefont {J.}~\bibnamefont {Bleuse}},
  \bibinfo {author} {\bibfnamefont {N.~S.}\ \bibnamefont {Malik}}, \bibinfo
  {author} {\bibfnamefont {M.}~\bibnamefont {Bazin}}, \bibinfo {author}
  {\bibfnamefont {P.}~\bibnamefont {Jaffrennou}}, \bibinfo {author}
  {\bibfnamefont {N.}~\bibnamefont {Gregersen}}, \bibinfo {author}
  {\bibfnamefont {C.}~\bibnamefont {Sauvan}}, \bibinfo {author} {\bibfnamefont
  {P.}~\bibnamefont {Lalanne}}, \ and\ \bibinfo {author} {\bibfnamefont
  {J.-M.}\ \bibnamefont {Gerard}},\ }\href
  {http://dx.doi.org/10.1038/nphoton.2009.287} {\bibfield  {journal} {\bibinfo
  {journal} {Nat Photon}\ }\textbf {\bibinfo {volume} {4}},\ \bibinfo {pages}
  {174} (\bibinfo {year} {2010})}\BibitemShut {NoStop}%
\bibitem [{\citenamefont {Schlehahn}\ \emph {et~al.}(2015)\citenamefont
  {Schlehahn}, \citenamefont {Gaafar}, \citenamefont {Vaupel}, \citenamefont
  {Gschrey}, \citenamefont {Schnauber}, \citenamefont {Schulze}, \citenamefont
  {Rodt}, \citenamefont {Strittmatter}, \citenamefont {Stolz}, \citenamefont
  {Rahimi-Iman}, \citenamefont {Heindel}, \citenamefont {Koch},\ and\
  \citenamefont {Reitzenstein}}]{143MHz:Reitzenstein}%
  \BibitemOpen
  \bibfield  {author} {\bibinfo {author} {\bibfnamefont {A.}~\bibnamefont
  {Schlehahn}}, \bibinfo {author} {\bibfnamefont {M.}~\bibnamefont {Gaafar}},
  \bibinfo {author} {\bibfnamefont {M.}~\bibnamefont {Vaupel}}, \bibinfo
  {author} {\bibfnamefont {M.}~\bibnamefont {Gschrey}}, \bibinfo {author}
  {\bibfnamefont {P.}~\bibnamefont {Schnauber}}, \bibinfo {author}
  {\bibfnamefont {J.-H.}\ \bibnamefont {Schulze}}, \bibinfo {author}
  {\bibfnamefont {S.}~\bibnamefont {Rodt}}, \bibinfo {author} {\bibfnamefont
  {A.}~\bibnamefont {Strittmatter}}, \bibinfo {author} {\bibfnamefont
  {W.}~\bibnamefont {Stolz}}, \bibinfo {author} {\bibfnamefont
  {A.}~\bibnamefont {Rahimi-Iman}}, \bibinfo {author} {\bibfnamefont
  {T.}~\bibnamefont {Heindel}}, \bibinfo {author} {\bibfnamefont
  {M.}~\bibnamefont {Koch}}, \ and\ \bibinfo {author} {\bibfnamefont
  {S.}~\bibnamefont {Reitzenstein}},\ }\href {\doibase
  http://dx.doi.org/10.1063/1.4927429} {\bibfield  {journal} {\bibinfo
  {journal} {Applied Physics Letters}\ }\textbf {\bibinfo {volume} {107}},\
  \bibinfo {eid} {041105} (\bibinfo {year} {2015}),\
  http://dx.doi.org/10.1063/1.4927429}\BibitemShut {NoStop}%
\bibitem [{\citenamefont {Strauf}\ \emph {et~al.}(2007)\citenamefont {Strauf},
  \citenamefont {Stoltz}, \citenamefont {Rakher}, \citenamefont {Coldren},
  \citenamefont {Petroff},\ and\ \citenamefont
  {Bouwmeester}}]{spQDHighFreq:Bouwm}%
  \BibitemOpen
  \bibfield  {author} {\bibinfo {author} {\bibfnamefont {S.}~\bibnamefont
  {Strauf}}, \bibinfo {author} {\bibfnamefont {N.~G.}\ \bibnamefont {Stoltz}},
  \bibinfo {author} {\bibfnamefont {M.~T.}\ \bibnamefont {Rakher}}, \bibinfo
  {author} {\bibfnamefont {L.~A.}\ \bibnamefont {Coldren}}, \bibinfo {author}
  {\bibfnamefont {P.~M.}\ \bibnamefont {Petroff}}, \ and\ \bibinfo {author}
  {\bibfnamefont {D.}~\bibnamefont {Bouwmeester}},\ }\href
  {http://dx.doi.org/10.1038/nphoton.2007.227} {\bibfield  {journal} {\bibinfo
  {journal} {Nat Photon}\ }\textbf {\bibinfo {volume} {1}},\ \bibinfo {pages}
  {704} (\bibinfo {year} {2007})}\BibitemShut {NoStop}%
\bibitem [{\citenamefont {Unsleber}\ \emph
  {et~al.}(2015{\natexlab{a}})\citenamefont {Unsleber}, \citenamefont
  {McCutcheon}, \citenamefont {Dambach}, \citenamefont {Lermer}, \citenamefont
  {Gregersen}, \citenamefont {H\"ofling}, \citenamefont {M\o{}rk},
  \citenamefont {Schneider},\ and\ \citenamefont {Kamp}}]{TPIDephasing:Kamp}%
  \BibitemOpen
  \bibfield  {author} {\bibinfo {author} {\bibfnamefont {S.}~\bibnamefont
  {Unsleber}}, \bibinfo {author} {\bibfnamefont {D.~P.~S.}\ \bibnamefont
  {McCutcheon}}, \bibinfo {author} {\bibfnamefont {M.}~\bibnamefont {Dambach}},
  \bibinfo {author} {\bibfnamefont {M.}~\bibnamefont {Lermer}}, \bibinfo
  {author} {\bibfnamefont {N.}~\bibnamefont {Gregersen}}, \bibinfo {author}
  {\bibfnamefont {S.}~\bibnamefont {H\"ofling}}, \bibinfo {author}
  {\bibfnamefont {J.}~\bibnamefont {M\o{}rk}}, \bibinfo {author} {\bibfnamefont
  {C.}~\bibnamefont {Schneider}}, \ and\ \bibinfo {author} {\bibfnamefont
  {M.}~\bibnamefont {Kamp}},\ }\href {\doibase 10.1103/PhysRevB.91.075413}
  {\bibfield  {journal} {\bibinfo  {journal} {Phys. Rev. B}\ }\textbf {\bibinfo
  {volume} {91}},\ \bibinfo {pages} {075413} (\bibinfo {year}
  {2015}{\natexlab{a}})}\BibitemShut {NoStop}%
\bibitem [{\citenamefont {Giesz}\ \emph
  {et~al.}(2015{\natexlab{a}})\citenamefont {Giesz}, \citenamefont {Portalupi},
  \citenamefont {Grange}, \citenamefont {Ant\'on}, \citenamefont {De~Santis},
  \citenamefont {Demory}, \citenamefont {Somaschi}, \citenamefont {Sagnes},
  \citenamefont {Lema\^{\i}tre}, \citenamefont {Lanco}, \citenamefont
  {Auff\`eves},\ and\ \citenamefont {Senellart}}]{cavEnhTPI:Giesz}%
  \BibitemOpen
  \bibfield  {author} {\bibinfo {author} {\bibfnamefont {V.}~\bibnamefont
  {Giesz}}, \bibinfo {author} {\bibfnamefont {S.~L.}\ \bibnamefont
  {Portalupi}}, \bibinfo {author} {\bibfnamefont {T.}~\bibnamefont {Grange}},
  \bibinfo {author} {\bibfnamefont {C.}~\bibnamefont {Ant\'on}}, \bibinfo
  {author} {\bibfnamefont {L.}~\bibnamefont {De~Santis}}, \bibinfo {author}
  {\bibfnamefont {J.}~\bibnamefont {Demory}}, \bibinfo {author} {\bibfnamefont
  {N.}~\bibnamefont {Somaschi}}, \bibinfo {author} {\bibfnamefont
  {I.}~\bibnamefont {Sagnes}}, \bibinfo {author} {\bibfnamefont
  {A.}~\bibnamefont {Lema\^{\i}tre}}, \bibinfo {author} {\bibfnamefont
  {L.}~\bibnamefont {Lanco}}, \bibinfo {author} {\bibfnamefont
  {A.}~\bibnamefont {Auff\`eves}}, \ and\ \bibinfo {author} {\bibfnamefont
  {P.}~\bibnamefont {Senellart}},\ }\href {\doibase 10.1103/PhysRevB.92.161302}
  {\bibfield  {journal} {\bibinfo  {journal} {Phys. Rev. B}\ }\textbf {\bibinfo
  {volume} {92}},\ \bibinfo {pages} {161302} (\bibinfo {year}
  {2015}{\natexlab{a}})}\BibitemShut {NoStop}%
\bibitem [{\citenamefont {Reimer}\ \emph {et~al.}(2014)\citenamefont {Reimer},
  \citenamefont {Bulgarini}, \citenamefont {Heeres}, \citenamefont {Witek},
  \citenamefont {Versteegh}, \citenamefont {Dalacu}, \citenamefont {J.},
  \citenamefont {Poole},\ and\ \citenamefont
  {Zwiller}}]{spectral-diff-nanowire}%
  \BibitemOpen
  \bibfield  {author} {\bibinfo {author} {\bibfnamefont {M.~E.}\ \bibnamefont
  {Reimer}}, \bibinfo {author} {\bibfnamefont {G.}~\bibnamefont {Bulgarini}},
  \bibinfo {author} {\bibfnamefont {R.~W.}\ \bibnamefont {Heeres}}, \bibinfo
  {author} {\bibfnamefont {B.~J.}\ \bibnamefont {Witek}}, \bibinfo {author}
  {\bibfnamefont {M.}~\bibnamefont {Versteegh}}, \bibinfo {author}
  {\bibfnamefont {D.}~\bibnamefont {Dalacu}}, \bibinfo {author} {\bibfnamefont
  {L.}~\bibnamefont {J.}}, \bibinfo {author} {\bibfnamefont {P.~J.}\
  \bibnamefont {Poole}}, \ and\ \bibinfo {author} {\bibfnamefont
  {V.}~\bibnamefont {Zwiller}},\ }\href@noop {} {\bibfield  {journal} {\bibinfo
   {journal} {arXiv:1407.2833}\ } (\bibinfo {year} {2014})}\BibitemShut
  {NoStop}%
\bibitem [{\citenamefont {Gazzano}\ \emph
  {et~al.}(2013{\natexlab{b}})\citenamefont {Gazzano}, \citenamefont {Almeida},
  \citenamefont {Nowak}, \citenamefont {Portalupi}, \citenamefont
  {Lema\^{i}tre}, \citenamefont {Sagnes}, \citenamefont {White},\ and\
  \citenamefont {Senellart}}]{entGate:Senellart}%
  \BibitemOpen
  \bibfield  {author} {\bibinfo {author} {\bibfnamefont {O.}~\bibnamefont
  {Gazzano}}, \bibinfo {author} {\bibfnamefont {M.~P.}\ \bibnamefont
  {Almeida}}, \bibinfo {author} {\bibfnamefont {A.~K.}\ \bibnamefont {Nowak}},
  \bibinfo {author} {\bibfnamefont {S.~L.}\ \bibnamefont {Portalupi}}, \bibinfo
  {author} {\bibfnamefont {A.}~\bibnamefont {Lema\^{i}tre}}, \bibinfo {author}
  {\bibfnamefont {I.}~\bibnamefont {Sagnes}}, \bibinfo {author} {\bibfnamefont
  {A.~G.}\ \bibnamefont {White}}, \ and\ \bibinfo {author} {\bibfnamefont
  {P.}~\bibnamefont {Senellart}},\ }\href {\doibase
  10.1103/PhysRevLett.110.250501} {\bibfield  {journal} {\bibinfo  {journal}
  {Phys. Rev. Lett.}\ }\textbf {\bibinfo {volume} {110}},\ \bibinfo {pages}
  {250501} (\bibinfo {year} {2013}{\natexlab{b}})}\BibitemShut {NoStop}%
\bibitem [{\citenamefont {Kuhlmann}\ \emph {et~al.}(2013)\citenamefont
  {Kuhlmann}, \citenamefont {Houel}, \citenamefont {Ludwig}, \citenamefont
  {Greuter}, \citenamefont {Reuter}, \citenamefont {Wieck}, \citenamefont
  {Poggio},\ and\ \citenamefont {Warburton}}]{indNoise:Warburton}%
  \BibitemOpen
  \bibfield  {author} {\bibinfo {author} {\bibfnamefont {A.~V.}\ \bibnamefont
  {Kuhlmann}}, \bibinfo {author} {\bibfnamefont {J.}~\bibnamefont {Houel}},
  \bibinfo {author} {\bibfnamefont {A.}~\bibnamefont {Ludwig}}, \bibinfo
  {author} {\bibfnamefont {L.}~\bibnamefont {Greuter}}, \bibinfo {author}
  {\bibfnamefont {D.}~\bibnamefont {Reuter}}, \bibinfo {author} {\bibfnamefont
  {A.~D.}\ \bibnamefont {Wieck}}, \bibinfo {author} {\bibfnamefont
  {M.}~\bibnamefont {Poggio}}, \ and\ \bibinfo {author} {\bibfnamefont {R.~J.}\
  \bibnamefont {Warburton}},\ }\href {http://dx.doi.org/10.1038/nphys2688}
  {\bibfield  {journal} {\bibinfo  {journal} {Nat Phys}\ }\textbf {\bibinfo
  {volume} {9}},\ \bibinfo {pages} {570} (\bibinfo {year} {2013})}\BibitemShut
  {NoStop}%
\bibitem [{\citenamefont {Kuhlmann}\ \emph {et~al.}(2015)\citenamefont
  {Kuhlmann}, \citenamefont {Prechtel}, \citenamefont {Houel}, \citenamefont
  {Ludwig}, \citenamefont {Reuter}, \citenamefont {Wieck},\ and\ \citenamefont
  {Warburton}}]{transLimitedQD:Warburton}%
  \BibitemOpen
  \bibfield  {author} {\bibinfo {author} {\bibfnamefont {A.~V.}\ \bibnamefont
  {Kuhlmann}}, \bibinfo {author} {\bibfnamefont {J.~H.}\ \bibnamefont
  {Prechtel}}, \bibinfo {author} {\bibfnamefont {J.}~\bibnamefont {Houel}},
  \bibinfo {author} {\bibfnamefont {A.}~\bibnamefont {Ludwig}}, \bibinfo
  {author} {\bibfnamefont {D.}~\bibnamefont {Reuter}}, \bibinfo {author}
  {\bibfnamefont {A.~D.}\ \bibnamefont {Wieck}}, \ and\ \bibinfo {author}
  {\bibfnamefont {R.~J.}\ \bibnamefont {Warburton}},\ }\href
  {http://dx.doi.org/10.1038/ncomms9204} {\bibfield  {journal} {\bibinfo
  {journal} {Nat Commun}\ }\textbf {\bibinfo {volume} {6}} (\bibinfo {year}
  {2015})}\BibitemShut {NoStop}%
\bibitem [{\citenamefont {Giesz}\ \emph
  {et~al.}(2015{\natexlab{b}})\citenamefont {Giesz}, \citenamefont {Somaschi},
  \citenamefont {Hornecker}, \citenamefont {Grange}, \citenamefont
  {Reznychenko}, \citenamefont {De~Santis}, \citenamefont {Demory},
  \citenamefont {Gomez}, \citenamefont {Sagnes}, \citenamefont {Lemaitre},
  \citenamefont {Krebs}, \citenamefont {Lanzillotti}, \citenamefont {Lanco},
  \citenamefont {Auffeves},\ and\ \citenamefont {Senellart}}]{fewPhoton:Giesz}%
  \BibitemOpen
  \bibfield  {author} {\bibinfo {author} {\bibfnamefont {V.}~\bibnamefont
  {Giesz}}, \bibinfo {author} {\bibfnamefont {N.}~\bibnamefont {Somaschi}},
  \bibinfo {author} {\bibfnamefont {G.}~\bibnamefont {Hornecker}}, \bibinfo
  {author} {\bibfnamefont {T.}~\bibnamefont {Grange}}, \bibinfo {author}
  {\bibfnamefont {B.}~\bibnamefont {Reznychenko}}, \bibinfo {author}
  {\bibfnamefont {L.}~\bibnamefont {De~Santis}}, \bibinfo {author}
  {\bibfnamefont {J.}~\bibnamefont {Demory}}, \bibinfo {author} {\bibfnamefont
  {C.}~\bibnamefont {Gomez}}, \bibinfo {author} {\bibfnamefont
  {I.}~\bibnamefont {Sagnes}}, \bibinfo {author} {\bibfnamefont
  {A.}~\bibnamefont {Lemaitre}}, \bibinfo {author} {\bibfnamefont
  {O.}~\bibnamefont {Krebs}}, \bibinfo {author} {\bibfnamefont {N.~D.}\
  \bibnamefont {Lanzillotti}}, \bibinfo {author} {\bibfnamefont
  {L.}~\bibnamefont {Lanco}}, \bibinfo {author} {\bibfnamefont
  {A.}~\bibnamefont {Auffeves}}, \ and\ \bibinfo {author} {\bibfnamefont
  {P.}~\bibnamefont {Senellart}},\ }\href@noop {} {\bibfield  {journal}
  {\bibinfo  {journal} {arXiv:1512.04725}\ } (\bibinfo {year}
  {2015}{\natexlab{b}})}\BibitemShut {NoStop}%
\bibitem [{\citenamefont {Humphreys}\ \emph {et~al.}(2013)\citenamefont
  {Humphreys}, \citenamefont {Metcalf}, \citenamefont {Spring}, \citenamefont
  {Moore}, \citenamefont {Jin}, \citenamefont {Barbieri}, \citenamefont
  {Kolthammer},\ and\ \citenamefont {Walmsley}}]{LOQCSPM:Walmsley}%
  \BibitemOpen
  \bibfield  {author} {\bibinfo {author} {\bibfnamefont {P.~C.}\ \bibnamefont
  {Humphreys}}, \bibinfo {author} {\bibfnamefont {B.~J.}\ \bibnamefont
  {Metcalf}}, \bibinfo {author} {\bibfnamefont {J.~B.}\ \bibnamefont {Spring}},
  \bibinfo {author} {\bibfnamefont {M.}~\bibnamefont {Moore}}, \bibinfo
  {author} {\bibfnamefont {X.-M.}\ \bibnamefont {Jin}}, \bibinfo {author}
  {\bibfnamefont {M.}~\bibnamefont {Barbieri}}, \bibinfo {author}
  {\bibfnamefont {W.~S.}\ \bibnamefont {Kolthammer}}, \ and\ \bibinfo {author}
  {\bibfnamefont {I.~A.}\ \bibnamefont {Walmsley}},\ }\href {\doibase
  10.1103/PhysRevLett.111.150501} {\bibfield  {journal} {\bibinfo  {journal}
  {Phys. Rev. Lett.}\ }\textbf {\bibinfo {volume} {111}},\ \bibinfo {pages}
  {150501} (\bibinfo {year} {2013})}\BibitemShut {NoStop}%
\bibitem [{\citenamefont {Rohde}(2015)}]{LOQCTLoop:Rohde}%
  \BibitemOpen
  \bibfield  {author} {\bibinfo {author} {\bibfnamefont {P.~P.}\ \bibnamefont
  {Rohde}},\ }\href {\doibase 10.1103/PhysRevA.91.012306} {\bibfield  {journal}
  {\bibinfo  {journal} {Phys. Rev. A}\ }\textbf {\bibinfo {volume} {91}},\
  \bibinfo {pages} {012306} (\bibinfo {year} {2015})}\BibitemShut {NoStop}%
\bibitem [{\citenamefont {Bentivegna}\ \emph {et~al.}(2015)\citenamefont
  {Bentivegna}, \citenamefont {Spagnolo}, \citenamefont {Vitelli},
  \citenamefont {Flamini}, \citenamefont {Viggianiello}, \citenamefont
  {Latmiral}, \citenamefont {Mataloni}, \citenamefont {Brod}, \citenamefont
  {Galv{\~a}o}, \citenamefont {Crespi}, \citenamefont {Ramponi}, \citenamefont
  {Osellame},\ and\ \citenamefont {Sciarrino}}]{ScattBS:Sciarrino}%
  \BibitemOpen
  \bibfield  {author} {\bibinfo {author} {\bibfnamefont {M.}~\bibnamefont
  {Bentivegna}}, \bibinfo {author} {\bibfnamefont {N.}~\bibnamefont
  {Spagnolo}}, \bibinfo {author} {\bibfnamefont {C.}~\bibnamefont {Vitelli}},
  \bibinfo {author} {\bibfnamefont {F.}~\bibnamefont {Flamini}}, \bibinfo
  {author} {\bibfnamefont {N.}~\bibnamefont {Viggianiello}}, \bibinfo {author}
  {\bibfnamefont {L.}~\bibnamefont {Latmiral}}, \bibinfo {author}
  {\bibfnamefont {P.}~\bibnamefont {Mataloni}}, \bibinfo {author}
  {\bibfnamefont {D.~J.}\ \bibnamefont {Brod}}, \bibinfo {author}
  {\bibfnamefont {E.~F.}\ \bibnamefont {Galv{\~a}o}}, \bibinfo {author}
  {\bibfnamefont {A.}~\bibnamefont {Crespi}}, \bibinfo {author} {\bibfnamefont
  {R.}~\bibnamefont {Ramponi}}, \bibinfo {author} {\bibfnamefont
  {R.}~\bibnamefont {Osellame}}, \ and\ \bibinfo {author} {\bibfnamefont
  {F.}~\bibnamefont {Sciarrino}},\ }\href {\doibase 10.1126/sciadv.1400255}
  {\bibfield  {journal} {\bibinfo  {journal} {Science Advances}\ }\textbf
  {\bibinfo {volume} {1}},\ \bibinfo {pages} {1400255} (\bibinfo {year}
  {2015})}\BibitemShut {NoStop}%
\bibitem [{\citenamefont {Zhang}\ \emph {et~al.}(2015)\citenamefont {Zhang},
  \citenamefont {Huang}, \citenamefont {Wang}, \citenamefont {Liu},
  \citenamefont {Li},\ and\ \citenamefont {Guo}}]{6photon:Guo}%
  \BibitemOpen
  \bibfield  {author} {\bibinfo {author} {\bibfnamefont {C.}~\bibnamefont
  {Zhang}}, \bibinfo {author} {\bibfnamefont {Y.-F.}\ \bibnamefont {Huang}},
  \bibinfo {author} {\bibfnamefont {Z.}~\bibnamefont {Wang}}, \bibinfo {author}
  {\bibfnamefont {B.-H.}\ \bibnamefont {Liu}}, \bibinfo {author} {\bibfnamefont
  {C.-F.}\ \bibnamefont {Li}}, \ and\ \bibinfo {author} {\bibfnamefont {G.-C.}\
  \bibnamefont {Guo}},\ }\href {\doibase 10.1103/PhysRevLett.115.260402}
  {\bibfield  {journal} {\bibinfo  {journal} {Phys. Rev. Lett.}\ }\textbf
  {\bibinfo {volume} {115}},\ \bibinfo {pages} {260402} (\bibinfo {year}
  {2015})}\BibitemShut {NoStop}%
\bibitem [{\citenamefont {Loredo}\ \emph {et~al.}(2016)\citenamefont {Loredo},
  \citenamefont {Broome}, \citenamefont {Hilaire}, \citenamefont {Gazzano},
  \citenamefont {Sagnes}, \citenamefont {Lemaitre}, \citenamefont {Almeida},
  \citenamefont {Senellart},\ and\ \citenamefont {White}}]{BS:Loredo}%
  \BibitemOpen
  \bibfield  {author} {\bibinfo {author} {\bibfnamefont {J.~C.}\ \bibnamefont
  {Loredo}}, \bibinfo {author} {\bibfnamefont {M.~A.}\ \bibnamefont {Broome}},
  \bibinfo {author} {\bibfnamefont {P.}~\bibnamefont {Hilaire}}, \bibinfo
  {author} {\bibfnamefont {O.}~\bibnamefont {Gazzano}}, \bibinfo {author}
  {\bibfnamefont {I.}~\bibnamefont {Sagnes}}, \bibinfo {author} {\bibfnamefont
  {A.}~\bibnamefont {Lemaitre}}, \bibinfo {author} {\bibfnamefont {M.~P.}\
  \bibnamefont {Almeida}}, \bibinfo {author} {\bibfnamefont {P.}~\bibnamefont
  {Senellart}}, \ and\ \bibinfo {author} {\bibfnamefont {A.~G.}\ \bibnamefont
  {White}},\ }\href@noop {} {\bibfield  {journal} {\bibinfo  {journal}
  {arXiv:1603.00054}\ } (\bibinfo {year} {2016})}\BibitemShut {NoStop}%
\bibitem [{\citenamefont {Unsleber}\ \emph
  {et~al.}(2015{\natexlab{b}})\citenamefont {Unsleber}, \citenamefont {He},
  \citenamefont {Maier}, \citenamefont {Gerhardt}, \citenamefont {Lu},
  \citenamefont {Pan}, \citenamefont {Kamp}, \citenamefont {Schneider},\ and\
  \citenamefont {Hofling}}]{detPillar75:Hofling}%
  \BibitemOpen
  \bibfield  {author} {\bibinfo {author} {\bibfnamefont {S.}~\bibnamefont
  {Unsleber}}, \bibinfo {author} {\bibfnamefont {Y.-M.}\ \bibnamefont {He}},
  \bibinfo {author} {\bibfnamefont {S.}~\bibnamefont {Maier}}, \bibinfo
  {author} {\bibfnamefont {S.}~\bibnamefont {Gerhardt}}, \bibinfo {author}
  {\bibfnamefont {C.-Y.}\ \bibnamefont {Lu}}, \bibinfo {author} {\bibfnamefont
  {J.-W.}\ \bibnamefont {Pan}}, \bibinfo {author} {\bibfnamefont
  {M.}~\bibnamefont {Kamp}}, \bibinfo {author} {\bibfnamefont {C.}~\bibnamefont
  {Schneider}}, \ and\ \bibinfo {author} {\bibfnamefont {S.}~\bibnamefont
  {Hofling}},\ }\href@noop {} {\bibfield  {journal} {\bibinfo  {journal}
  {arXiv:1512.07453v1}\ } (\bibinfo {year} {2015}{\natexlab{b}})}\BibitemShut
  {NoStop}%
\end{thebibliography}%

\newpage
\section{Supplementary Material}

\subsection{Areas in time-correlation histograms}

\begin{figure*}[htp]
\centering
\includegraphics[width=.95\linewidth]{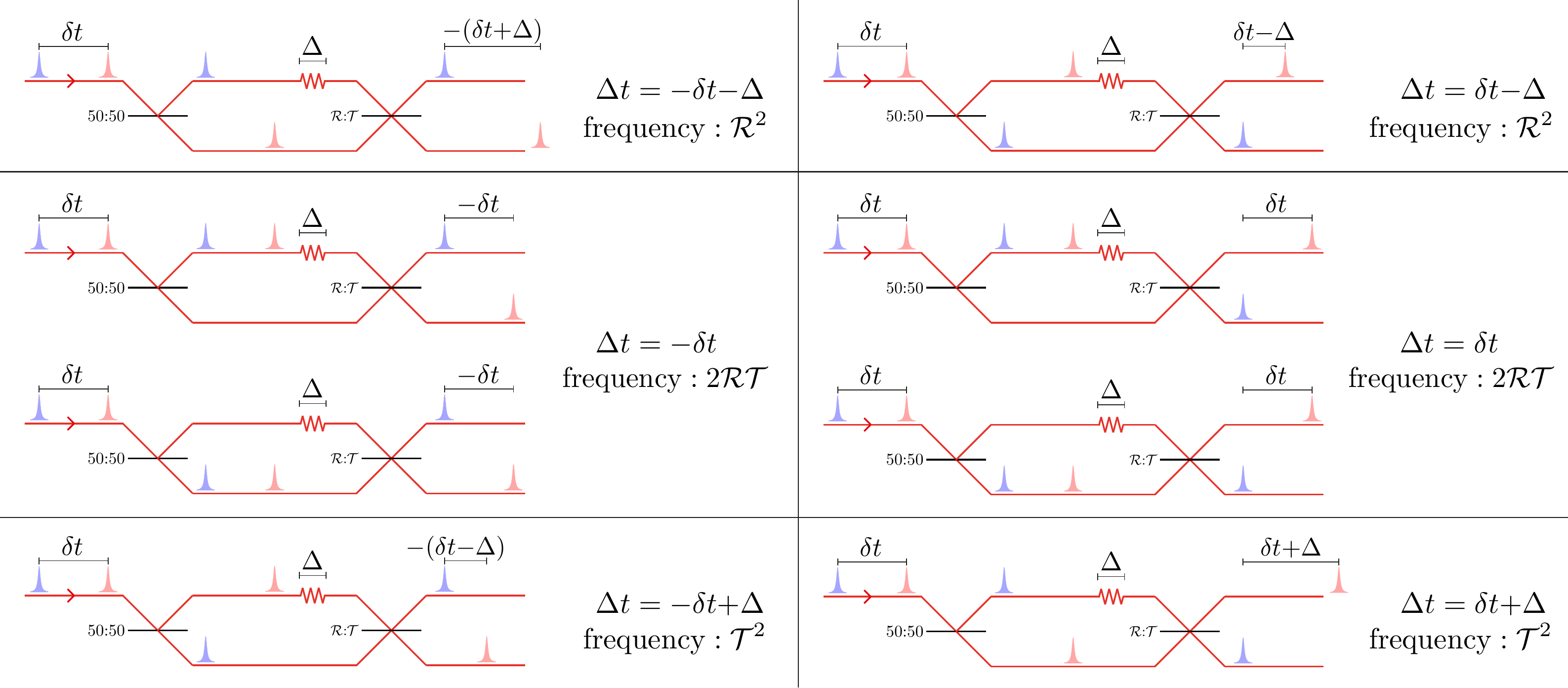}
\caption{Two consecutive single-photons separated by $\delta t$ passing through a $\Delta$-unbalanced Mach-Zehnder interferometer. $8$ outcome distributions, occurring with a given relative frequency, lead to a coincidence signal between events separated in time by $\Delta t$. The relative delay $\Delta t$ is positive if a detector in the upper output fires first, and it is negative in the opposite case.}
\label{fig:s1}
\end{figure*}
Here we deduce the area distribution of the time-correlation measurements described in the main text. For simplicity, we first consider two (fully-distinguishable) single-photons distributed in time-bins $\{t_1,t_2\}$, entering an unbalanced Mach-Zehnder interferometer composed of a first $50{:}50$ beamsplitter and a second beamsplitter with reflectance $\mathcal{R}$ (transmittance $\mathcal{T}{=}1{-}\mathcal{R}$). Our task is to find all possible output distributions leading to a coincidence detection between events separated in time by $\Delta t$. There are two timescales relevant in such coincidence measurements: the difference in occupied time-bins $\delta t{=}|t_2{-}t_1|$, and the temporal delay inside the unbalanced interferometer $\Delta$. By inspecting this reduced scenario, we can find that there are $8$ events leading to a coincidence detection, as depicted in Fig.~\ref{fig:s1}. This results in local patterns of peak areas $A_{\Delta t}$ given by: $A_{{-}\delta t{-}\Delta}{=}\mathcal{R}^2$, $A_{{-}\delta t}{=}2\mathcal{R}\mathcal{T}$, and $A_{{-}\delta t{+}\Delta}{=}\mathcal{T}^2$, the local pattern around ${-}\delta t$; and $A_{\delta t{-}\Delta}{=}\mathcal{R}^2$, $A_{\delta t}{=}2\mathcal{R}\mathcal{T}$, and $A_{\delta t{+}\Delta}{=}\mathcal{T}^2$, the local pattern around $\delta t$. From this, we find simple rules for the time-correlation measurement of an array of single-photons distributed in arbitrary time-bins $\{t_i\}$ passing through a $\Delta$-unbalanced Mach-Zehnder:

\emph{rule~1}:~Find all possible temporal delays $\delta t$ relating each pair of photons within the given time-bin distribution.

\emph{rule~2}:~Around each $\pm\delta t$, assign the relative frequency of events $\{\mathcal{R}^2,2\mathcal{R}\mathcal{T},\mathcal{T}^2\}$ at temporal delays $\Delta t{=}\{\pm\delta t{-}\Delta,\pm\delta t,\pm\delta t{+}\Delta\}$.

We note that these two simple rules describe different interesting histograms relevant in the literature. For instance, by simply identifying the involved parameters, one can find histograms of $g^{(2)}(\Delta t)$ measurements of arbitrary $|n\rangle$ Fock states by considering $n$ single-photons occupying the same time-bin, resulting in distributions agreeing with $g^{(2)}(0){=}1{-}1{/}n$, or the well known $5$-peak structures in two-photon interference experiments involving pairs of photons separated by $\Delta\tau_e<12.5$~ns repeated every $12.5$~ns.

Now, the experiment described in the main text is the particular case of an infinitely long stream of single-photons separated by a fixed $\delta t{=}12.5$~ns, and passing through an unbalanced interferometer with $\Delta{=}\Delta\tau_e$. Under this consideration, and following \emph{rule~1} and \emph{rule~2}, we derive the distribution of areas $A_{\Delta t}$, given by: $A_{k}{=}N$, $A_{-\Delta\tau_e}{=}N(1{-}\mathcal{R}^2)$, $A_{\Delta\tau_e}{=}N(1{-}\mathcal{T}^2)$, and $A_0{=}N\left(\left(\mathcal{R}^2+\mathcal{T}^2\right)-2\mathcal{R}\mathcal{T}V\right)$, with $k{=}\pm12.5\text{~ns},\pm25\text{~ns},...$, excluding peaks at $\pm\Delta\tau_e$, and $N$ an integration constant. The visibility term $V$ in $A_0$ appears from noticing (in virtue of \emph{rule~1} and \emph{rule~2}) that the area at $\Delta t{=}0$ for fully-distinguishable photons is $A_0^{V{=}0}{=}N(\mathcal{R}^2+\mathcal{T}^2)$, and then one simply uses the well-known relation $V{=}\left( 1{-} A_0{/}A_0^{V{=}0}\right)\left( \mathcal{R}^2{+}\mathcal{T}^2 \right){/}\left( 2\mathcal{R}\mathcal{T} \right)$, with $A_0$ relating the coincidence rate at zero delay of photons with non-zero $V$ indistinguishability. 

\subsection{Visibility power-dependence}

\begin{figure}[h]
\centering
\includegraphics[width=.85\linewidth]{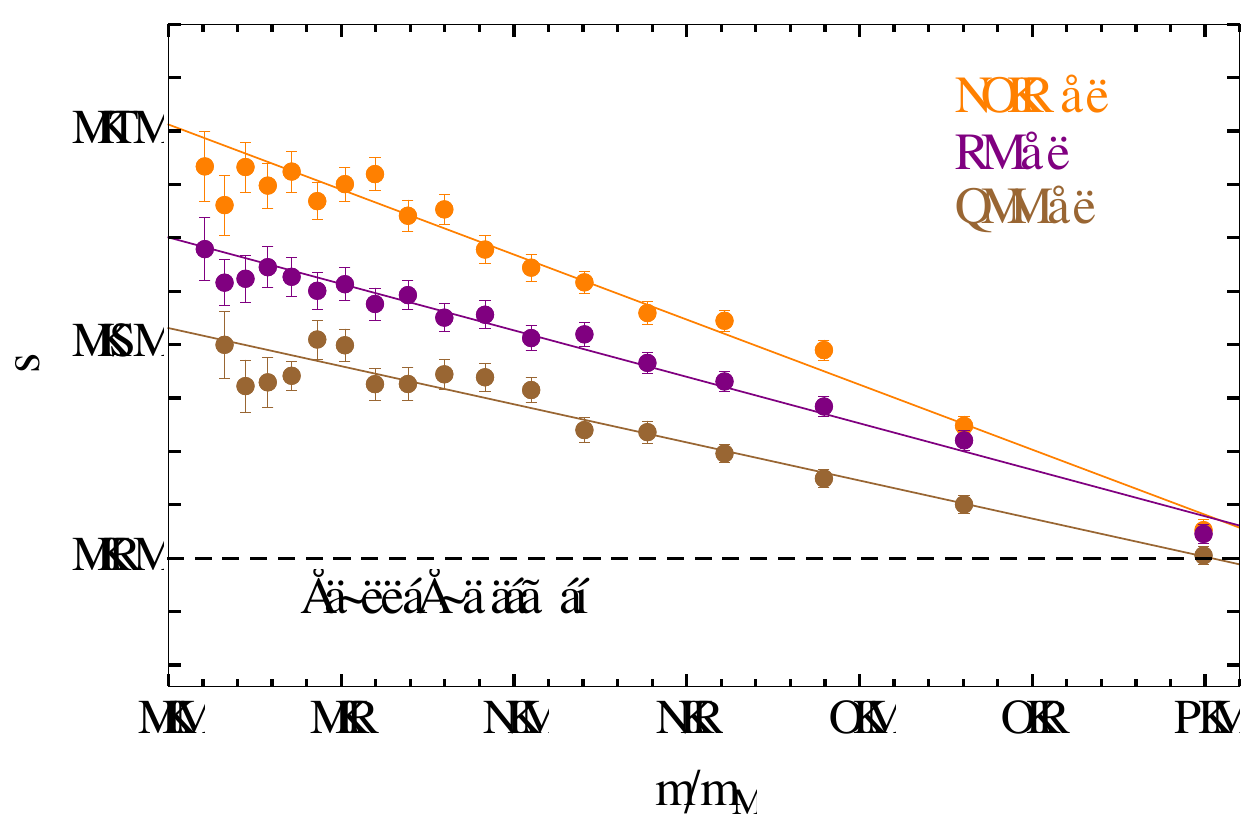}
\caption{Power-dependence of $V$ for $\Delta\tau_e{=}12.5$~ns (orange), $\Delta\tau_e{=}50$~ns (purple), and $\Delta\tau_e{=}400$~ns (brown). Curves are fits to $\overline{V}{=}V_{\Delta\tau_e}^{max}{+}m_{\Delta\tau_e}P$. $V$ is above $50\%$ (the classical limit) at all powers and timescales here explored.}
\label{fig:s2}
\end{figure}
Following the main text, the interference visibility $V$ of two photons separated in time by $\Delta\tau_e$ exhibits a linear-dependence in the pump power $P$. For a given $\Delta\tau_e$, we measure $V$ at various values of $P$, up to three saturation powers $P{=}3P_0$, and fit the data to $\overline{V}{=}V_{\Delta\tau_e}^{max}{+}m_{\Delta\tau_e}P$. Figure~{\ref{fig:s2}} shows the power-dependence of $V$ for $\Delta\tau_e{=}12.5$~ns, $\Delta\tau_e{=}50$~ns, and $\Delta\tau_e{=}400$~ns. The fitted parameters are $V_{12.5\text{ns}}^{max}{=}(70.3{\pm}0.3)\%$, $m_{12.5\text{ns}}{=}{-}(6.1{\pm}0.2)\%$ at short timescales; $V_{50\text{ns}}^{max}{=}(65.0{\pm}0.3)\%$, $m_{50\text{ns}}{=}{-}(4.4{\pm}0.2)\%$ at moderate timescales; and $V_{400\text{ns}}^{max}{=}(60.8{\pm}0.3)\%$, $m_{400\text{ns}}{=}{-}(3.6{\pm}0.2)\%$ at the longest timescales explored in this work.

\subsection{Visibilities of temporally-distant photons}

The interference visibility of two photons from two sources $a$ and $b$ reads~\cite{cavEnhTPI:Giesz}:
	\begin{equation}\label{eq:s1}
		V= \left(\frac{\gamma_a \gamma_b}{\gamma_a+\gamma_b}\right) \frac{ (\gamma_a+\gamma_b+\gamma^*_a+\gamma^*_b) } {\left[(\gamma_a+\gamma_b+\gamma^*_a+\gamma^*_b)/2\right]^2+\delta \omega^2},
	\end{equation}
where the $\gamma_i$ are the radiative rates, $\gamma^*_i$ the pure dephasing rates, and $\delta \omega$ the frequency detuning between the two sources. If the interfering photons are emitted by the same quantum dot, we assume that $\gamma_a{=}\gamma_b{=}\gamma$ and $\gamma^*_a{=}\gamma^*_b{=}\gamma^*$ are constant, but only the frequency $\omega{=}\omega_0+\delta\omega(t)$ varies over time (i.e. spectral wandering) around a central value $\omega_0$. This model makes sense here as the timescale over which $\omega$ varies is much larger than the radiative lifetime. Then Eq.~(\ref{eq:s1}) reduces to:
	\begin{equation}\label{eq:s2}	
		V=\left<\frac{V(0)}{1+\delta\omega_r^2}\right>,
	\end{equation}
where we have used $V(0){=}\gamma{/}(\gamma+\gamma^*)$ the "intrinsic" degree of indistinguishability, and $\delta\omega_r{=}\delta\omega{/}(\gamma+\gamma^*)$ the ratio between the frequency detuning and the spectral linewidth $\gamma+\gamma^*$.

\begin{figure*}[t]
\centering
\includegraphics[width=.85\linewidth]{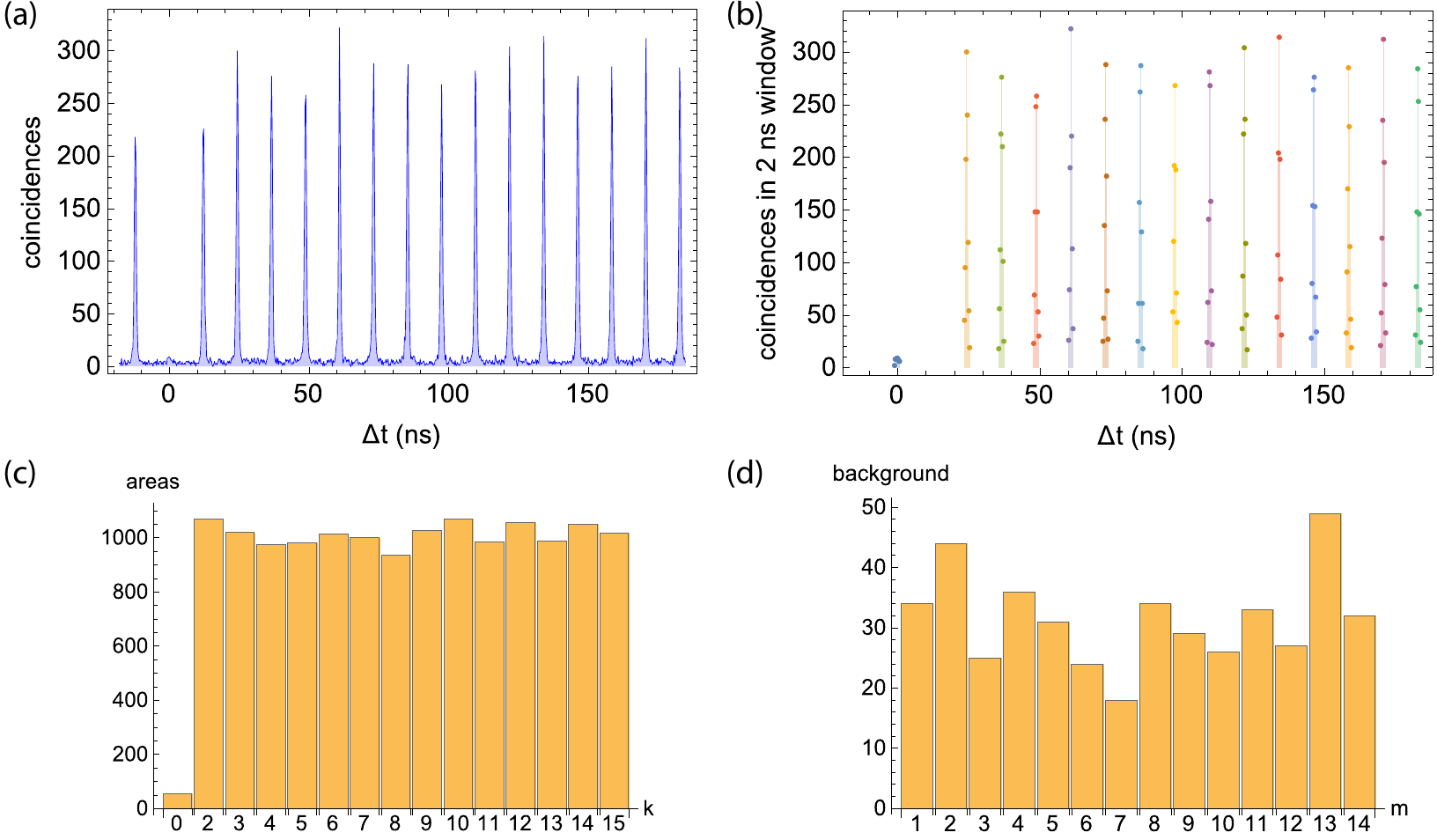}
\caption{Method to extract the {raw} and {corrected} interference visibilities. a) Interference histogram of two photons separated by $\Delta\tau_e{=}12.2$~ns. b) Subset of data involved in the evaluation of $V$. c) Integrated counts from data in b). d) Measured background in between peaks.}
\label{fig:s3}
\end{figure*}
One can define a time correlation function for the frequency fluctuations as
	\begin{equation}
		F(\Delta\tau_e) = <\delta\omega(t)\delta\omega(t+\Delta\tau_e)> = <\delta \omega^2> f(\Delta\tau_e),
	\end{equation}
then, the frequency difference as a function of the delay $\Delta\tau_e$ can be expressed as
	\begin{eqnarray}\nonumber
		<\delta \omega^2(\Delta\tau_e)>&=& <(\delta\omega(t+\Delta\tau_e)-\delta\omega(t))^2>\\
		&=& 2 <\delta \omega^2> (1-f(\Delta\tau_e)).
	\end{eqnarray}
A common assumption is to assume an exponential correlation function
\begin{equation}
 f(\Delta\tau_e) = e^{-\Delta\tau_e / \tau_c},
\end{equation}
with $\tau_c$ a characteristic wandering timescale. Which is expected for a Markovian dynamics of the environnement. An additional input which is required is the distribution for $\delta\omega$. Generally one assumes a Gaussian distribution, but for simplicity, and without loss of generality, we take a two-value distribution $\delta\omega = \pm \sqrt{ <\delta \omega^2>} $, so that: 
	\begin{eqnarray}\nonumber
		V(\Delta\tau_e) &=&\left<\frac{V(0)}{1+\delta\omega_r^2(\Delta\tau_e)}\right>\\ \nonumber
&=& \frac{V(0)}{1+\left<\delta\omega_r^2(\Delta\tau_e)\right>} \\
&=& \frac{V(0)}{1+2\delta\omega_r^2\left(1-e^{-\Delta\tau_e/\tau_c}\right)}
	\end{eqnarray}

\subsection{Extraction of visibility under resonant excitation}

\begin{figure}[h]
\centering
\includegraphics[width=.85\linewidth]{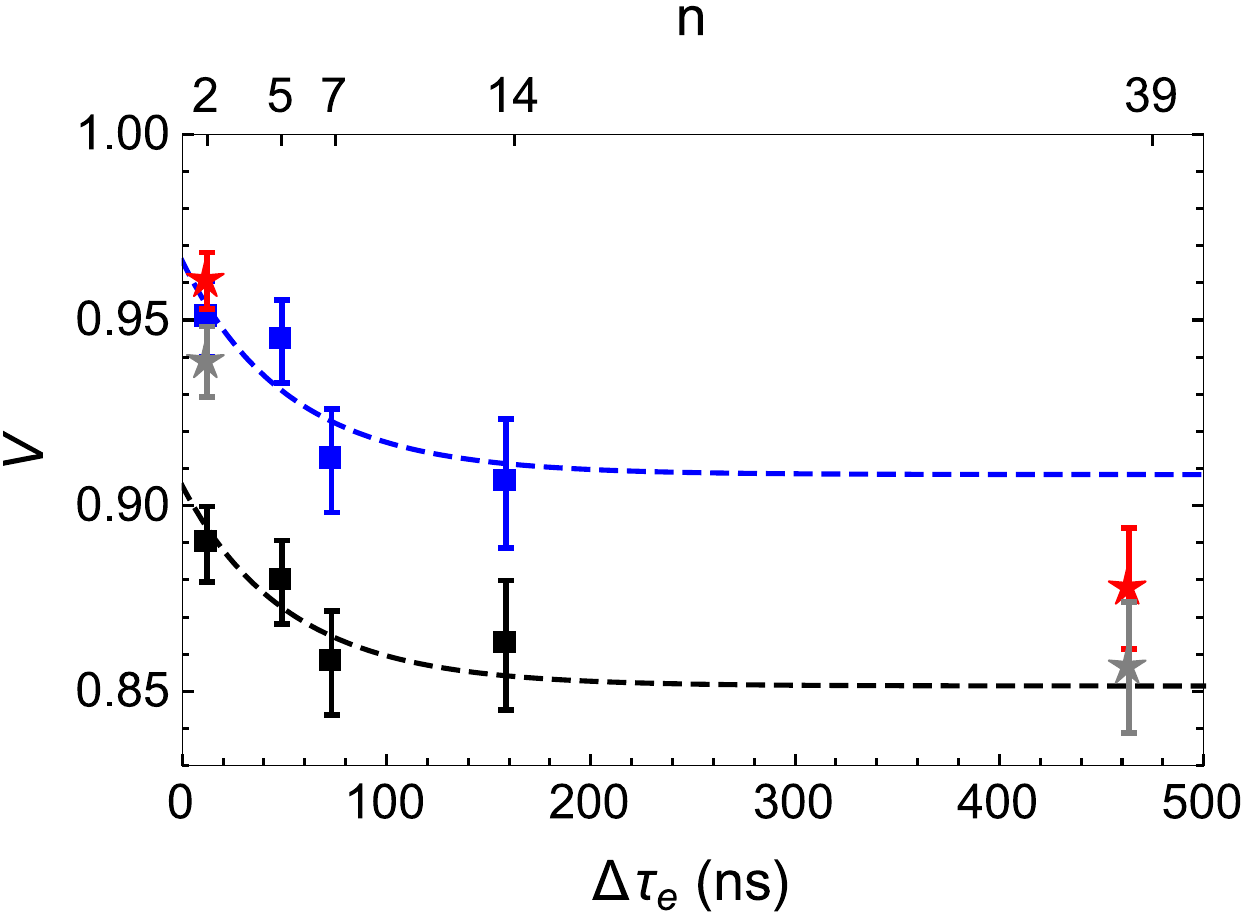}
\caption{{Indistinguishability vs temporal distance. Blue squares are corrected indistinguishabilities taken with \emph{Device~2}, and red stars are the corrected values taken with \emph{Device~3}. Black squares are raw values from \emph{Device~2}, and gray stars are raw values from \emph{Device~3}.}}
\label{fig:s4}
\end{figure}
Here we describe the methods to extract the {raw} and {corrected} two-photon interference visibilities under strictly-resonant excitation and $\pi$-pulse preparation, see Fig.~\ref{fig:s3}. Figure~\ref{fig:s3}a shows the interference histogram of two photons separated by $\Delta\tau_e{=}12.2$~ns, from which a visibility is extracted via $V{=}\left(\mathcal{R}^2+\mathcal{T}^2{-}A_0{/}A\right)/\left(2\mathcal{R}\mathcal{T}\right)$, where $A_0$ is the area of the peak around $\Delta t{=}0$, and $A$ is taken as the average area of $14$ adjacent peaks (excluding the peak at $\Delta\tau_e$). These areas are taken as the integrated counts within a temporal window of $2$~ns (considerably longer than the subnanosecond lifetimes) around $\Delta t{=}k\times12.2$~ns, with $k{=}0,2,3,...,15$, see Fig.~\ref{fig:s3}b. The resulting integrated areas are shown in Fig.~\ref{fig:s3}c, from which we extract a {raw} $V^{\pi}_{12.2\text{ns}}{=}(89.0\pm1.5)\%$. As described in the main text, the remaining non-vanishing area at $\Delta t{=}0$ is indeed quite small and it is on the order of experimental noise. We take into account this noise by integrating coincidence counts within a $2$~ns window but now located in between peaks, that is at $\Delta t{=}(m+1/2)\times12.2$~ns, with $m{=}1,2,...,14$, see Fig.\ref{fig:s3}d. After subtracting the average of these background counts to the areas in Fig.\ref{fig:s3}c, we obtained the {corrected} visibility $V^{\pi}_{12.2\text{ns}}{=}(95.0\pm1.0)\%$. These same methods were employed for all measurements under strictly-resonant excitation. {Figure~\ref{fig:s4} shows both raw and corrected visibilities for two devices (\emph{Device~2} and 3) extracted with this method.}

Measurements under quasi-resonant excitation, as described in the main text, exhibit a noise level $<1\%$, and therefore no noise-correction was employed.

\end{document}